\documentclass[12pt]{article}
\usepackage{amssymb}
\usepackage{amsmath}
\usepackage{graphicx}
\usepackage{indentfirst}
\usepackage{cite}
\usepackage{appendix}

\allowdisplaybreaks

\begin{document}

\title{Production of $\psi (4040)$, $\psi (4160)$, and $\psi (4415)$ mesons in
hadronic matter}
\author{Li-Yuan Li$^1$, Xiao-Ming Xu$^1$, and H. J. Weber$^2$}
\date{}
\maketitle \vspace{-1cm}
\centerline{$^1$Department of Physics, Shanghai University, Baoshan,
Shanghai 200444, China}
\centerline{$^2$Department of Physics, University of Virginia, Charlottesville,
VA 22904, USA}

\begin{abstract}
We present the first study of production of $\psi (4040)$, $\psi (4160)$,
and $\psi (4415)$ mesons in hadronic matter. Quark interchange between two
colliding charmed mesons leads to the production of the three mesons.
We calculate unpolarized cross sections for the reactions, 
$D\bar{D} \to \rho R$, $D\bar{D}^* \to \pi R$, $D\bar{D}^* \to \rho R$, 
$D^*\bar{D} \to \pi R$, $D^*\bar{D} \to \rho R$,
$D^*\bar{D}^* \to \pi R$, and $D^*\bar{D}^* \to \rho R$, where $R$ stands for
$\psi (4040)$, $\psi (4160)$, and $\psi (4415)$. In the temperature region
covering hadronic matter the peak cross sections of producing 
$\psi (4040)$ are similar to or larger than the ones of producing 
$\psi (4415)$, and the latter are generally larger than those of producing
$\psi (4160)$. With the cross sections we establish new master rate equations
for $\psi (4040)$, $\psi (4160)$, and $\psi (4415)$. The equations
are solved for central Pb-Pb collisions at $\sqrt{s_{NN}}=5.02$ TeV at the 
Large Hadron Collider.
Solutions of the equations show that the $\psi (4040)$ number density at
kinetic freeze-out of hadronic matter is larger than the $\psi (4415)$ number
density which is larger than the $\psi (4160)$ number density.
\end{abstract}



\vspace{0.5cm}
\leftline{\bf I. INTRODUCTION}
\vspace{0.5cm}

The masses, decay widths, and decay modes of
$\psi (4040)$, $\psi (4160)$, and $\psi (4415)$ mesons produced in
$e^+ e^-$ collisions have been measured by the Mark I Collaboration 
\cite{MARKI}, the DASP Collaboration 
\cite{DASP}, the BES Collaboration \cite{BES}, the
CLEO Collaboration \cite{CLEO}, the Belle Collaboration \cite{BELLE}, the
BABAR Collaboration \cite{BABAR}, and the BES III Collaboration \cite{BESIII}.
The three charmonia have also been studied in the $B^+ \to K^+ \mu^+ \mu^-$
decay and the
$B^+ \to D^+ D^- K^+$ decay by the LHCb Collaboration \cite{LHCb}.
In meson degrees of freedom, an electron-positron collision produces one
photon, and the photon then converts into $\psi (4040)$, $\psi (4160)$,
or $\psi (4415)$ \cite{PGK,BIO}. In quark degrees of freedom the photon 
produced
in the electron-positron annihilation splits into a charm quark and a charm
antiquark which evolve nonperturbatively into $\psi (4040)$, $\psi (4160)$, 
or $\psi (4415)$ with certain probabilities.

It is shown in Refs. \cite{GI,BGS,VFV} that $\psi (4040)$, $\psi (4160)$, and
$\psi (4415)$ can be individually interpreted as the $3 ^3S_1$,
$2 ^3D_1$, and $4 ^3S_1$ quark-antiquark states. In ultrarelativistic heavy-ion
collisions the three states created in initial nucleus-nucleus collisions
are dissolved in quark-gluon plasmas \cite{JXW}. At the critical temperature,
hadronization of
the quark-gluon plasma gives hadronic matter that evolves until
kinetic freeze-out. All possible hadron-hadron scattering occurs in hadronic
matter, and various quark processes appear during scattering \cite{XW}. Quark
interchange between two colliding charmed mesons in hadronic matter leads to
production of $\psi (4040)$, $\psi (4160)$, and $\psi (4415)$, for example,
$D\bar{D}^* \to \rho \psi (4040)$, $\rho \psi (4160)$, and
$\rho \psi (4415)$. Therefore, the three charmonia can be
produced in hadronic matter.

In the present work, we study the production of $\psi (4040)$, $\psi (4160)$, 
and $\psi (4415)$ via quark interchange between two colliding charmed 
mesons in hadronic matter. Since the temperature of hadronic matter varies 
during its expansion, 
we need to study temperature dependence of cross sections for 
the production of the three charmonia. Furthermore, we establish new master
rate equations to account for charmonium number densities that result from
the reactions of charmed mesons. Number densities are obtained from the
equations for central Pb-Pb collisions at the center-of-mass energy per
nucleon-nucleon pair $\sqrt{s_{NN}}=5.02$ TeV at the Large Hadron Collider
(LHC).

This paper is organized as follows. In Sect.~II we provide cross-section
formulas for 2-to-2 scattering of charmed mesons. In Sect.~III 
we give master rate equations for $\psi (4040)$, $\psi (4160)$, and 
$\psi (4415)$ mesons. In Sect.~IV we show numerical cross sections and 
number densities of the three mesons produced in hadronic matter. 
Relevant discussions are given. In Sect.~V we summarize the present work.

\vspace{0.5cm}
\leftline{\bf II. FORMALISM FOR $c\bar c$ PRODUCTION IN MESON-MESON}
\leftline{\bf COLLISIONS}
\vspace{0.5cm}
Production of $\psi(4040)$, $\psi(4160)$, and $\psi(4415)$ mesons in hadronic
matter concerns the reaction $A(c\bar{q}_{2})+B(q_{1}\bar{c})\rightarrow
C(q_{1}\bar{q}_{2})+D(c\bar{c})$
where $c$ is a charm quark, $q_{1}$ a light quark, and $\bar{q}_{2}$
a light antiquark. This reaction is caused by interchange of $c$
and $q_{1}$ between mesons $A$ and $B$. Denote the spin of meson
$A$ ($B$, $C$, $D$) by $S_{A}$ ($S_{B}$, $S_{C}$, $S_{D}$) and its magnetic
projection quantum number by $S_{Az}$ ($S_{Bz}$, $S_{Cz}$, $S_{Dz}$). The wave
function of mesons $A$ and $B$ is
\begin{equation}
\psi_{AB}=\phi_{A{\rm rel}}\phi_{B{\rm rel}}\phi_{A{\rm color}}
\phi_{B{\rm color}}\chi_{S_{A}S_{Az}}\chi_{S_{B}S_{Bz}}
\varphi_{AB{\rm flavor}},
\end{equation}
and the wave function of mesons $C$ and $D$ is
\begin{equation}
\psi_{CD}=\phi_{C{\rm rel}}\phi_{D{\rm rel}}\phi_{C{\rm color}}
\phi_{D{\rm color}}\chi_{S_{C}S_{Cz}}\chi_{S_{D}S_{Dz}}
\varphi_{CD{\rm flavor}},
\end{equation}
where $\phi_{A{\rm rel}}$ ($\phi_{B{\rm rel}}$, $\phi_{C{\rm rel}}$,
$\phi_{D{\rm rel}}$), $\phi_{A{\rm color}}$ ($\phi_{B{\rm color}}$,
$\phi_{C{\rm color}}$, $\phi_{D{\rm color}}$),
and $\chi_{S_{A}S_{Az}}$ ($\chi_{S_{B}S_{Bz}}$, $\chi_{S_{C}S_{Cz}}$,
$\chi_{S_{D}S_{Dz}}$)
are the mesonic quark-antiquark relative-motion wave function, the color wave
function, and the spin wave function of meson $A$ ($B$, $C$, $D$),
respectively; the flavor wave function $\varphi_{AB{\rm flavor}}$ of mesons $A$
and $B$ possesses the same isospin as the flavor wave function
$\varphi_{CD{\rm flavor}}$ of mesons $C$ and $D$.

Two forms are involved in the Born-order meson-meson scattering in the
quark interchange mechanism \cite{BS}.
Scattering in the prior form as seen in Fig. 1 means that gluon exchange
takes place prior to quark interchange. Scattering in the post form
shown  in Fig. 2 means quark interchange is followed by gluon exchange.
In the two figures the interaction between constituents $a$ and $b$ is
indicated by the dot-dashed line, and its potential is $V_{ab}$.
Scattering in the prior form needs interactions between the two constituents
of meson $A$ and the ones of meson $B$, which are described by the potentials
$V_{c\bar c}$, $V_{\bar{q}_2q_1}$, $V_{cq_1}$, and $V_{\bar{q}_2\bar c}$.
The transition amplitude for scattering in the prior form is
\begin{eqnarray}
{\cal M}_{\rm{fi}}^{\rm{prior}} &=&\sqrt{2E_A 2E_B 2E_C 2E_D}
\int\frac{d^{3}p_{q_{1}\bar{q}_{2}}}{(2\pi)^{3}}
\int\frac{d^{3}p_{c\bar{c}}}{(2\pi)^{3}}\psi_{CD}^{+}
\nonumber\\
& & [V_{c\bar{c}}(\vec{p}_{c}^{~\prime}-\vec{p}_{c})+V_{\bar{q}_{2}q_{1}}
(\vec{p}_{\bar{q}_{2}}^{~\prime}-\vec{p}_{\bar{q}_{2}})
+V_{cq_{1}}(\vec{p}_{c}^{~\prime}-\vec{p}_{c})+V_{\bar{q}_{2}\bar{c}}
(\vec{p}_{\bar{q}_{2}}^{~\prime}-\vec{p}_{\bar{q}_{2}})]\psi_{AB},
\end{eqnarray}
where $E_A$ ($E_B$, $E_C$, $E_D$) denotes the energy of meson $A$ ($B$, $C$,
$D$), and $\vec{p}_{ab}$ is the relative momentum of constituents $a$ and $b$.
The three-dimensional momentum of constituent $a$ in the initial (final) mesons
is labeled as $\vec{p}_a$ ($\vec{p}_a^{~\prime}$). 
Substituting Eqs. (1) and (2) in Eq. (3), we obtain
\begin{eqnarray}
{\cal M_{{\rm fi}}^{{\rm prior}}} & = &\sqrt{2E_A 2E_B 2E_C 2E_D}
\varphi_{CD{\rm flavor}}^{+}\varphi_{AB{\rm flavor}}\phi_{C{\rm color}}^{+}
\phi_{D{\rm color}}^{+}\chi_{S_{C}S_{Cz}}^{+}\chi_{S_{D}S_{Dz}}^{+}\nonumber\\
& & [\int\frac{d^{3}p_{c}}{(2\pi)^{3}}\frac{d^{3}p_{c}^{\prime}}{(2\pi)^{3}}
\phi_{C{\rm rel}}^{+}(\vec{p}_{c}-\vec{P}
+\frac{m_{\bar{q}_{2}}}{m_{q_{1}}+m_{\bar{q}_{2}}}\vec{P}^{\prime})
\phi_{D{\rm rel}}^{+}(\vec{p}_{c}^{~\prime}
+\frac{m_{c}}{m_{c}+m_{\bar{c}}}\vec{P}^{\prime})\nonumber\\
& & V_{c\bar{c}}(\vec{p}_{c}^{~\prime}-\vec{p}_{c})\phi_{A{\rm rel}}
(\vec{p}_{c}-\frac{m_{c}}{m_{c}+m_{\bar{q}_{2}}}\vec{P})
\phi_{B{\rm rel}}(\vec{p}_{c}
-\frac{m_{\bar{c}}}{m_{q_{1}}+m_{\bar{c}}}\vec{P}+\vec{P}^{\prime})\nonumber\\
& & +\int\frac{d^{3}p_{\bar{q}_{2}}}{(2\pi)^{3}}
\frac{d^{3}p_{\bar{q}_{2}}^{~\prime}}{(2\pi)^{3}}
\phi_{C{\rm rel}}^{+}(-\vec{p}_{\bar{q}_{2}}^{~\prime}
+\frac{m_{\bar{q}_{2}}}{m_{q_{1}}+m_{\bar{q}_{2}}}\vec{P}^{\prime})
\phi_{D{\rm rel}}^{+}(-\vec{p}_{\bar{q}_{2}}+\vec{P}
+\frac{m_{c}}{m_{c}+m_{\bar{c}}}\vec{P}^{\prime})\nonumber\\
& & V_{\bar{q}_{2}q_{1}}(\vec{p}_{\bar{q}_{2}}^{~\prime}-\vec{p}_{\bar{q}_{2}})
\phi_{A{\rm rel}}(-\vec{p}_{\bar{q}_{2}}
+\frac{m_{\bar{q}_{2}}}{m_{c}+m_{\bar{q}_{2}}}\vec{P})
\phi_{B{\rm rel}}(-\vec{p}_{\bar{q}_{2}}
+\frac{m_{q_{1}}}{m_{q_{1}}+m_{\bar{c}}}\vec{P}+\vec{P}^{\prime})\nonumber\\
& & +\int\frac{d^{3}p_{c}}{(2\pi)^{3}}\frac{d^{3}p_{c}^{\prime}}{(2\pi)^{3}}
\phi_{C{\rm rel}}^{+}(\vec{p}_{c}-\vec{P}
+\frac{m_{\bar{q}_{2}}}{m_{q_{1}}+m_{\bar{q}_{2}}}\vec{P}^{\prime})
\phi_{D{\rm rel}}^{+}(\vec{p}_{c}^{~\prime}
+\frac{m_{c}}{m_{c}+m_{\bar{c}}}\vec{P}^{\prime})\nonumber\\
& & V_{cq_{1}}(\vec{p}_{c}^{~\prime}-\vec{p}_{c})\phi_{A{\rm rel}}
(\vec{p}_{c}-\frac{m_{c}}{m_{c}+m_{\bar{q}_{2}}}\vec{P})
\phi_{B{\rm rel}}(\vec{p}_{c}^{~\prime}
-\frac{m_{\bar{c}}}{m_{q_{1}}+m_{\bar{c}}}\vec{P}+\vec{P}^{\prime})\nonumber\\
& & +\int\frac{d^{3}p_{\bar{q}_{2}}}{(2\pi)^{3}}
\frac{d^{3}p_{\bar{q}_{2}}^{\prime}}{(2\pi)^{3}}
\phi_{C{\rm rel}}^{+}(-\vec{p}_{\bar{q}_{2}}^{~\prime}
+\frac{m_{\bar{q}_{2}}}{m_{q_{1}}+m_{\bar{q}_{2}}}\vec{P}^{\prime})
\phi_{D{\rm rel}}^{+}(-\vec{p}_{\bar{q}_{2}}+\vec{P}
+\frac{m_{c}}{m_{c}+m_{\bar{c}}}\vec{P}^{\prime})\nonumber\\
& & V_{\bar{q}_{2}\bar{c}}(\vec{p}_{\bar{q}_{2}}^{~\prime}
-\vec{p}_{\bar{q}_{2}})
\phi_{A{\rm rel}}(-\vec{p}_{\bar{q}_{2}}
+\frac{m_{\bar{q}_{2}}}{m_{c}+m_{\bar{q}_{2}}}\vec{P})
\phi_{B{\rm rel}}(-\vec{p}_{\bar{q}_{2}}^{~\prime}
+\frac{m_{q_{1}}}{m_{q_{1}}+m_{\bar{c}}}\vec{P}+\vec{P}^{\prime})]\nonumber\\
& & \chi_{S_{A}S_{Az}}\chi_{S_{B}S_{Bz}}\phi_{A{\rm color}}\phi_{B{\rm color}},
\end{eqnarray}
where $m_a$ is the mass of constituent $a$;
$\vec{P}$ and $\vec{P}^{'}$ are the three-dimensional momenta of mesons A
and C in the center-of-mass frame of the two initial mesons, respectively.
The first, second, third, and fourth terms enclosed by the brackets correspond
to the upper left, upper right, lower left, and lower right diagrams 
in Fig. 1, respectively. 
Scattering in the post form needs interactions between the two constituents
of meson $C$ and the ones of meson $D$, which are described by the potentials
$V_{q_1\bar c}$, $V_{\bar{q}_2c}$, $V_{cq_1}$, and $V_{\bar{q}_2\bar c}$.
The transition amplitude for scattering in the post form is
\begin{eqnarray}
{\cal M}_{\rm{fi}}^{\rm{post}} &=&\sqrt{2E_A 2E_B 2E_C 2E_D}
\int\frac{d^{3}p_{c\bar{q}_{2}}}{(2\pi)^{3}}
\int\frac{d^{3}p_{q_{1}\bar{c}}}{(2\pi)^{3}}\psi_{CD}^{+}\nonumber\\
& & [V_{q_{1}\bar{c}}(\vec{p}_{q_{1}}^{~\prime}-\vec{p}_{q_{1}})+
V_{\bar{q}_{2}c}(\vec{p}_{\bar{q}_{2}}^{~\prime}-\vec{p}_{\bar{q}_{2}})
+V_{cq_{1}}(\vec{p}_{c}^{~\prime}-\vec{p}_{c})+V_{\bar{q}_{2}\bar{c}}
(\vec{p}_{\bar{q}_{2}}^{~\prime}-\vec{p}_{\bar{q}_{2}})]
\psi_{AB}.
\end{eqnarray}
With the wave functions of the initial mesons and the final mesons, we have
\begin{eqnarray}
{\cal M_{{\rm fi}}^{{\rm post}}} & = &\sqrt{2E_A 2E_B 2E_C 2E_D}
\varphi_{CD{\rm flavor}}^{+}
\varphi_{AB{\rm flavor}}\phi_{C{\rm color}}^{+}
\phi_{D{\rm color}}^{+}\chi_{S_{C}S_{Cz}}^{+}\chi_{S_{D}S_{Dz}}^{+}\nonumber\\
& & [\int\frac{d^{3}p_{q_{1}}}{(2\pi)^{3}}
\frac{d^{3}p_{q_{1}}^{\prime}}{(2\pi)^{3}}
\phi_{C{\rm rel}}^{+}(\vec{p}_{q_{1}}^{~\prime}
-\frac{m_{q_{1}}}{m_{q_{1}}+m_{\bar{q}_{2}}}\vec{P}^{\prime})
\phi_{D{\rm rel}}^{+}(\vec{p}_{q_{1}}^{~\prime}+\vec{P}
-\frac{m_{\bar{c}}}{m_{c}+m_{\bar{c}}}\vec{P}^{\prime})\nonumber\\
& & V_{q_{1}\bar{c}}(\vec{p}_{q_{1}}^{~\prime}-\vec{p}_{q_{1}})
\phi_{A{\rm rel}}(\vec{p}_{q_{1}}^{~\prime}
+\frac{m_{\bar{q}_{2}}}{m_{c}+m_{\bar{q}_{2}}}\vec{P}-\vec{P}^{\prime})
\phi_{B{\rm rel}}(\vec{p}_{q_{1}}
+\frac{m_{q_{1}}}{m_{q_{1}}+m_{\bar{c}}}\vec{P})\nonumber\\
& & +\int\frac{d^{3}p_{\bar{q}_{2}}}{(2\pi)^{3}}
\frac{d^{3}p_{\bar{q}_{2}}^{\prime}}{(2\pi)^{3}}
\phi_{C{\rm rel}}^{+}(-\vec{p}_{\bar{q}_{2}}^{~\prime}
+\frac{m_{\bar{q}_{2}}}{m_{q_{1}}+m_{\bar{q}_{2}}}\vec{P}^{\prime})
\phi_{D{\rm rel}}^{+}(-\vec{p}_{\bar{q}_{2}}^{~\prime}+\vec{P}
+\frac{m_{c}}{m_{c}+m_{\bar{c}}}\vec{P}^{\prime})\nonumber\\
& & V_{\bar{q}_{2}c}(\vec{p}_{\bar{q}_{2}}^{~\prime}-\vec{p}_{\bar{q}_{2}})
\phi_{A{\rm rel}}(-\vec{p}_{\bar{q}_{2}}
+\frac{m_{\bar{q}_{2}}}{m_{c}+m_{\bar{q}_{2}}}\vec{P})
\phi_{B{\rm rel}}(-\vec{p}_{\bar{q}_{2}}^{~\prime}
+\frac{m_{q_{1}}}{m_{q_{1}}+m_{\bar{c}}}\vec{P}+\vec{P}^{\prime})\nonumber\\
& & +\int\frac{d^{3}p_{c}}{(2\pi)^{3}}\frac{d^{3}p_{c}^{\prime}}{(2\pi)^{3}}
\phi_{C{\rm rel}}^{+}(\vec{p}_{c}-\vec{P}
+\frac{m_{\bar{q}_{2}}}{m_{q_{1}}+m_{\bar{q}_{2}}}\vec{P}^{\prime})
\phi_{D{\rm rel}}^{+}(\vec{p}_{c}^{~\prime}
+\frac{m_{c}}{m_{c}+m_{\bar{c}}}\vec{P}^{\prime})\nonumber\\
& & V_{cq_{1}}(\vec{p}_{c}^{~\prime}-\vec{p}_{c})\phi_{A{\rm rel}}
(\vec{p}_{c}-\frac{m_{c}}{m_{c}+m_{\bar{q}_{2}}}\vec{P})
\phi_{B{\rm rel}}(\vec{p}_{c}^{~\prime}
-\frac{m_{\bar{c}}}{m_{q_{1}}+m_{\bar{c}}}\vec{P}+\vec{P}^{\prime})\nonumber\\
& & +\int\frac{d^{3}p_{\bar{q}_{2}}}{(2\pi)^{3}}
\frac{d^{3}p_{\bar{q}_{2}}^{\prime}}{(2\pi)^{3}}
\phi_{C{\rm rel}}^{+}(-\vec{p}_{\bar{q}_{2}}^{~\prime}
+\frac{m_{\bar{q}_{2}}}{m_{q_{1}}+m_{\bar{q}_{2}}}\vec{P}^{\prime})
\phi_{D{\rm rel}}^{+}(-\vec{p}_{\bar{q}_{2}}+\vec{P}
+\frac{m_{c}}{m_{c}+m_{\bar{c}}}\vec{P}^{\prime})\nonumber\\
& & V_{\bar{q}_{2}\bar{c}}(\vec{p}_{\bar{q}_{2}}^{~\prime}
-\vec{p}_{\bar{q}_{2}})
\phi_{A{\rm rel}}(-\vec{p}_{\bar{q}_{2}}
+\frac{m_{\bar{q}_{2}}}{m_{c}+m_{\bar{q}_{2}}}\vec{P})
\phi_{B{\rm rel}}(-\vec{p}_{\bar{q}_{2}}^{~\prime}
+\frac{m_{q_{1}}}{m_{q_{1}}+m_{\bar{c}}}\vec{P}+\vec{P}^{\prime})]\nonumber\\
& & \chi_{S_{A}S_{Az}}\chi_{S_{B}S_{Bz}}\phi_{A{\rm color}}\phi_{B{\rm color}}.
\end{eqnarray}
The first, second, third, and fourth terms enclosed by the brackets correspond
to the upper left, upper right, lower left, and lower right diagrams 
in Fig. 2, respectively. 
In Eqs. (4) and (6), $\phi_{A{\rm rel}}$, $\phi_{B{\rm rel}}$, 
$\phi_{C{\rm rel}}$,
and $\phi_{D{\rm rel}}$ are functions of $\vec{p}_{c\bar{q}_{2}}$,  
$\vec{p}_{q_{1}\bar{c}}$, $\vec{p}_{q_{1}\bar{q}_{2}}$,
and $\vec{p}_{c\bar{c}}$, respectively. These relative momenta
equal the expressions enclosed by the parentheses that follow
$\phi_{A{\rm rel}}$, $\phi_{B{\rm rel}}$, $\phi_{C{\rm rel}}$, 
and $\phi_{D{\rm rel}}$. In Eqs. (3)-(6), $V_{c\bar c}$, $V_{\bar{q}_2q_1}$,
$V_{cq_1}$, $V_{\bar{q}_2\bar c}$, $V_{q_1\bar c}$, and $V_{\bar{q}_2c}$ are
momentum-space potentials that depend on momenta attached to the dot-dashed
lines in Figs. 1 and 2. The momenta equal the expressions enclosed by the 
parentheses that follow $V_{c\bar c}$, $V_{\bar{q}_2q_1}$,
$V_{cq_1}$, $V_{\bar{q}_2\bar c}$, $V_{q_1\bar c}$, and $V_{\bar{q}_2c}$.

As seen in Eqs. (4) and (6), each of $\cal M_{\rm fi}^{\rm prior}$ and
$\cal M_{\rm fi}^{\rm post}$ contains four integrals. The first and second 
terms of $\cal M_{\rm fi}^{\rm prior}$ look different than the first and 
second terms of $\cal M_{\rm fi}^{\rm post}$, but the third and fourth terms
of $\cal M_{\rm fi}^{\rm prior}$ are identical with the third and fourth terms
of $\cal M_{\rm fi}^{\rm post}$. If the sum of the first and second terms of
$\cal M_{\rm fi}^{\rm prior}$ does not equal the one of 
$\cal M_{\rm fi}^{\rm post}$, $\cal M_{\rm fi}^{\rm prior}$ does not
equal $\cal M_{\rm fi}^{\rm post}$, and the post-prior ambiguity appears.

Denote by $T$ the temperature and by $T_{\rm c}$ the critical
temperature between quark-gluon plasmas and hadronic matter. Let $\vec {s}_a$
be the spin of constituent $a$. 
The mesonic quark-antiquark relative-motion wave functions, $\phi_{A\rm rel}$,
$\phi_{B\rm rel}$, $\phi_{C\rm rel}$, and $\phi_{D\rm rel}$, are the Fourier 
transform of the solutions of the Schr\"odinger equation with the
potential between constituents $a$ and $b$ in coordinate space \cite{LXW},
\begin{eqnarray}
V_{ab}(\vec{r}_{ab}) & = &
- \frac {\vec{\lambda}_a}{2} \cdot \frac {\vec{\lambda}_b}{2}
\xi_1 \left[ 1.3- \left( \frac {T}{T_{\rm c}} \right)^4 \right] \tanh
(\xi_2 r_{ab}) + \frac {\vec{\lambda}_a}{2} \cdot \frac {\vec{\lambda}_b}{2}
\frac {6\pi}{25} \frac {v(\lambda r_{ab})}{r_{ab}} \exp (-\xi_3 r_{ab})
\nonumber  \\
& & -\frac {\vec{\lambda}_a}{2} \cdot \frac {\vec{\lambda}_b}{2}
\frac {16\pi^2}{25}\frac{d^3}{\pi^{3/2}}\exp(-d^2r^2_{ab})
\frac {\vec {s}_a \cdot \vec {s}_b} {m_am_b}
+\frac {\vec{\lambda}_a}{2} \cdot \frac {\vec{\lambda}_b}{2}\frac {4\pi}{25}
\frac {1} {r_{ab}} \frac {d^2v(\lambda r_{ab})}{dr_{ab}^2}
\frac {\vec {s}_a \cdot \vec {s}_b}{m_am_b}
\nonumber  \\
& & -\frac {\vec{\lambda}_a}{2} \cdot \frac {\vec{\lambda}_b}{2}
\frac {6\pi}{25m_am_b}\left[ v(\lambda r_{ab})
-r_{ab}\frac {dv(\lambda r_{ab})}{dr_{ab}} +\frac{r_{ab}^2}{3}
\frac {d^2v(\lambda r_{ab})}{dr_{ab}^2} \right]
\nonumber  \\
& & \left( \frac{3\vec {s}_a \cdot \vec{r}_{ab}\vec {s}_b \cdot \vec{r}_{ab}}
{r_{ab}^5} -\frac {\vec {s}_a \cdot \vec {s}_b}{r_{ab}^3} \right) ,
\end{eqnarray}
where $\vec{r}_{ab}$ is the relative coordinate of constituents $a$ and $b$;
$\xi_1=0.525$ GeV, $\xi_2=1.5[0.75+0.25 (T/{T_{\rm c}})^{10}]^6$ GeV, 
$\xi_3=0.6$ GeV, $T_{\rm c}=0.175$ GeV, and
$\lambda=\sqrt{25/16\pi^2 \alpha'}$ with $\alpha'=1.04$ GeV$^{-2}$;
$v$ is a function of $r_{ab}$, and is given by Buchm\"uller and Tye in Ref. 
\cite{BT}; the quantity $d$ is related to constituent masses by
\begin{eqnarray}
d^2=d_1^2\left[\frac{1}{2}+\frac{1}{2}\left(\frac{4m_a m_b}{(m_a+m_b)^2}
\right)^4\right]+d_2^2\left(\frac{2m_am_b}{m_a+m_b}\right)^2,
\label{eq:d}
\end{eqnarray}
where $d_1=0.15$ GeV and $d_2=0.705$.
The short-distance part of the potential is obtained from perturbative 
quantum chromodynamics \cite{BT}, and the temperature dependence from
lattice gauge calculations \cite{KLP}. The lattice calculations gave a 
temperature-dependent quark potential at intermediate and long distances.
The potential at long distances has a distance-independent value that 
decreases with increasing temperature. The first term in Eq. (7) is the 
confining potential that corresponds to the lattice results. The expression
$\frac {\vec{\lambda}_a}{2} \cdot \frac {\vec{\lambda}_b}{2}
\frac {6\pi}{25} \frac {v(\lambda r_{ab})}{r_{ab}}$ in the second term
arises from one-gluon exchange plus perturbative one- and two-loop corrections
in vacuum \cite{BT}, and the factor $\exp (-\xi_3 r_{ab})$ is a medium
modification factor. The third term is the smeared spin-spin interaction that
comes from one-gluon exchange between constituents $a$ and $b$ \cite{GI}. The
fourth term is the spin-spin interaction that originates from perturbative 
one- and two-loop corrections to one-gluon exchange \cite{Xu2002}. The fifth
term is the tensor interaction that arises from one-gluon exchange plus 
perturbative one- and two-loop corrections \cite{Xu2002}.

In Fig. 3 we plot the central spin-independent potential which is the sum of 
the first and second terms on the right-hand side of Eq. (7), the spin-spin 
interaction which is the sum of the third and fourth terms, 
the tensor interaction, and the potential $V_{c\bar c}$ between the charm 
quark and the charm antiquark inside the $\psi(4160)$ meson at zero temperature
by the dotted, dashed, dot-dashed, and solid curves, respectively. The tensor
interaction dominates $V_{c\bar c}$ at very short distances,
but is short-range. At short distances, $V_{c\bar c}$ increases with
increasing distance, while the spin-spin interaction is positive and decreases
rapidly. At long distances, $V_{c\bar c}$ approaches 0.91 GeV, while
the spin-spin interaction almost becomes zero.

The Schr\"odinger equation with the potential $V_{ab} (\vec{r}_{ab})$
at zero temperature gives meson masses that are close to
the experimental masses of $\pi$, $\rho$, $K$, $K^*$, $J/\psi$, $\chi_{c}$, $
\psi'$, $\psi (3770)$, $\psi (4040)$, $\psi (4160)$, $\psi (4415)$, $D$, $D^*$,
$D_s$, and $D^*_s$
mesons~\cite{PDG}. The experimental data of $S$- and $P$-wave
elastic phase shifts for $\pi \pi$ scattering in
vacuum~\cite{pipiqi,pipianni} are reproduced in the Born approximation
\cite{JSX,SXW}. In the calculations of the meson masses and the phase shifts,
the masses of the up quark, the down quark, the strange quark, and the charm 
quark are kept as 0.32 GeV, 0.32 GeV, 0.5 GeV, and 1.51 GeV, respectively.

In Eqs. (4) and (6), the flavor matrix elements $\varphi_{CD{\rm flavor}}^{+}
\varphi_{AB{\rm flavor}}$ equal 1. Related to the potential, 
$\phi_{C{\rm color}}^{+} \phi_{D{\rm color}}^{+}\phi_{A{\rm color}}
\phi_{B{\rm color}}$ is $\frac{1}{3}$, and
$\phi_{C{\rm color}}^{+} \phi_{D{\rm color}}^{+}\frac {\vec{\lambda}_a}{2} 
\cdot \frac {\vec{\lambda}_b}{2}\phi_{A{\rm color}} \phi_{B{\rm color}}$
are $-\frac{4}{9}$, $-\frac{4}{9}$, $\frac{4}{9}$, $\frac{4}{9}$, 
$-\frac{4}{9}$, and $-\frac{4}{9}$ for $V_{c\bar c}$, $V_{\bar{q}_2 q_1}$, 
$V_{cq_1}$, $V_{\bar{q}_2 \bar{c}}$, $V_{q_1\bar c}$, and $V_{\bar{q}_2 c}$,
respectively.

From $m_A$, $m_B$, $m_C$, and $m_D$, which are the masses of mesons $A$, 
$B$, $C$, and $D$, $\vec P$ and $\vec{P}'$  are given by
\begin{equation}
{\vec P}^2(\sqrt{s})=\frac{1}{4s}[(s-m_A^2-m_B^2)^2-4m_A^2m_B^2],
\end{equation}
\begin{equation}
{\vec P}^{\prime 2}(\sqrt{s})=\frac{1}{4s}[(s-m_C^2-m_D^2)^2-4m_C^2m_D^2],
\end{equation}
where the Mandelstam variable $s=(P_A+P_B)^2$ is defined from the four-momenta
of mesons $A$ and $B$, $P_A$ and $P_B$.
Let $J_{Az}$ ($J_{Bz}$, $J_{Cz}$, $J_{Dz}$) denote the magnetic projection
quantum number of the total angular momentum $J_A$ ($J_B$, $J_C$, $J_D$) of 
meson $A$ ($B$, $C$, $D$). 
When the transition amplitude for scattering in the prior form equals the one
for scattering in the post form, i.e., no post-prior discrepancy exists, the
unpolarized cross section may be calculated from $\cal M_{\rm fi}^{\rm prior}$,
\begin{equation}
\sigma_{\rm unpol}^{\rm prior}(\sqrt {s},T) = \frac {1}{(2J_A+1)(2J_B+1)}
\frac{1}{32\pi s}\frac{|\vec{P}^{\prime }(\sqrt{s})|}{|\vec{P}(\sqrt{s})|}
\int_{0}^{\pi }d\theta \sum\limits_{J_{Az}J_{Bz}J_{Cz}J_{Dz}}
\mid {\cal M}_{\rm fi}^{\rm prior} \mid^2 \sin \theta ,
\end{equation}
or from $\cal M_{\rm fi}^{\rm post}$,
\begin{equation}
\sigma_{\rm unpol}^{\rm post}(\sqrt {s},T) = \frac {1}{(2J_A+1)(2J_B+1)}
\frac{1}{32\pi s}\frac{|\vec{P}^{\prime }(\sqrt{s})|}{|\vec{P}(\sqrt{s})|}
\int_{0}^{\pi }d\theta \sum\limits_{J_{Az}J_{Bz}J_{Cz}J_{Dz}}
\mid {\cal M}_{\rm fi}^{\rm post} \mid^2 \sin \theta ,
\end{equation}
where $\theta$ is the angle between $\vec{P}$ and $\vec{P}'$.
When the transition amplitude for scattering in the prior form does not equal
the one for scattering in the post form, i.e., the post-prior discrepancy 
occurs, we treat $\cal M_{\rm fi}^{\rm prior}$ and $\cal M_{\rm fi}^{\rm post}$
on a completely equal footing, and the unpolarized cross section for 
$A+B \to C+D$ is 
\begin{eqnarray}
\sigma^{\rm unpol}(\sqrt {s},T) & = & \frac{1}{2}
[\sigma_{\rm unpol}^{\rm prior}(\sqrt {s},T)
+\sigma_{\rm unpol}^{\rm post}(\sqrt {s},T)]
              \nonumber    \\
& = & \frac {1}{(2J_A+1)(2J_B+1)}
\frac{1}{64\pi s}\frac{|\vec{P}^{\prime }(\sqrt{s})|}{|\vec{P}(\sqrt{s})|}
              \nonumber    \\
& & \int_{0}^{\pi }d\theta \sum\limits_{J_{Az}J_{Bz}J_{Cz}J_{Dz}}
(\mid {\cal M}_{\rm fi}^{\rm prior} \mid^2 
+ \mid {\cal M}_{\rm fi}^{\rm post} \mid^2) \sin \theta .
\end{eqnarray}
In fact, this equation gives 
$\sigma^{\rm unpol}=\sigma_{\rm unpol}^{\rm prior}
=\sigma_{\rm unpol}^{\rm post}$ when 
${\cal M}_{\rm fi}^{\rm prior}={\cal M}_{\rm fi}^{\rm post}$. Hence, Eq. (13)
is used to calculate the unpolarized cross section. The 
cross section depends on temperature and the total energy of the two
initial mesons in the center-of-mass frame.

\vspace{0.5cm}
\leftline{\bf III. MASTER RATE EQUATIONS}
\vspace{0.5cm}

We use the notation
$D=\left( \begin{array}{c}D^+\\ D^0 \end{array} \right)$
and $\bar{D}=\left( \begin{array}{c}\bar{D}^0\\ D^- \end{array} \right)$ for
the pseudoscalar isospin doublets as well as
$D^{\ast}=\left( \begin{array}{c}D^{\ast +}\\ D^{\ast 0} \end{array} \right)$
and $\bar{D}^{\ast}=\left( \begin{array}{c}\bar{D}^{\ast 0}\\ D^{\ast -}
\end{array} \right)$ for the vector isospin doublets. In hadronic matter
number densities for $\psi(4040)$, $\psi(4160)$, and $\psi(4415)$ mesons change
with respect to time and space according to the following rate equations,
\begin{equation} \label{rate}
\partial_{\mu}(n_{R}u^{\mu})=\varTheta_{R},
\end{equation}
where $\mu$ is the space-time index, and
$u^{\mu}=(u^0,\vec {u})$ is the four-velocity of a fluid element in hadronic
matter.
$n_{\psi (4040)}$, $n_{\psi (4160)}$, and $n_{\psi (4415)}$ are the number
densities of $\psi (4040)$, $\psi (4160)$, and $\psi (4415)$, if $R$ stands for
$\psi (4040)$, $\psi (4160)$, and $\psi (4415)$, respectively.
In the present work, we take into account only these reactions,
\begin{displaymath}
D \bar{D} \to \rho R;~D \bar{D}^* \to \pi R,~\rho R;
~D^* \bar{D} \to \pi R, \rho R;~D^* \bar{D}^* \to \pi R, \rho R,
\end{displaymath}
where light mesons in these final states are limited to pions and $\rho$ 
mesons. The source terms are given by
\begin{align}
\varTheta_R=
&\langle\sigma_{D\bar{D} \rightarrow \rho R}v_{D\bar D}\rangle n_Dn_{\bar D}
+\langle\sigma_{D\bar{D}^* \rightarrow \pi R}v_{D\bar{D}^*}\rangle n_D
n_{\bar{D}^*}
 \notag \\
&+\langle\sigma_{D^*\bar{D} \rightarrow \pi R}v_{D^*\bar D}\rangle n_{D^*}
n_{\bar D}
+\langle\sigma_{D\bar{D}^* \rightarrow \rho R}v_{D\bar{D}^*}\rangle n_D
n_{\bar{D}^*} \notag \\
&+\langle\sigma_{D^*\bar{D} \rightarrow \rho R}v_{D^*\bar D}\rangle n_{D^*}
n_{\bar D}
+\langle\sigma_{D^*\bar{D}^* \rightarrow \pi R}v_{D^*\bar{D}^*}\rangle n_{D^*}
n_{\bar{D}^*}
\notag \\
&+\langle\sigma_{D^*\bar{D}^* \rightarrow \rho R}v_{D^*\bar{D}^*}\rangle
n_{D^*}n_{\bar{D}^*}. \notag \\
\end{align}
The thermal-averaged cross section with
the relative velocity of two initial mesons is defined as
\begin{equation}
\langle \sigma_{ij\to i^\prime j^\prime} v_{ij}\rangle
=\frac{\int\frac{d^3 k_i}{(2\pi)^3}f_i(k_i)
\frac{d^3 k_j}{(2\pi)^3}f_j(k_j)\sigma_{ij \to i^\prime j^\prime}
(\sqrt{s}, T)v_{ij}}
{\int\frac{d^3 k_i}{(2\pi)^3}f_i(k_i)\int\frac{d^3 k_j}{(2\pi)^3}
f_j(k_j)},
\end{equation}
where $f_i(k_i)$ and $f_j(k_j)$ are the momentum distribution functions of
mesons $i$ and $j$ with the four-momenta $k_i$ and $k_j$ in the rest frame of
hadronic matter, respectively;
$\sigma_{ij \to i^\prime j^\prime}(\sqrt{s})$ is the isospin-averaged
unpolarized cross section for $ij\to i^\prime j^\prime$, and is obtained from
the unpolarized cross section given in Eq. (13) according to formulas given in
Appendix A; $v_{ij}$ is the relative velocity of meson $i$ 
with mass $m_i$ and meson $j$ with mass $m_j$, 
\begin{equation}
v_{ij}=\frac{\sqrt{(k_i\cdot k_j)^2-m_i^2 m_j^2}}{k_i^0 k_j^0}. 
\end{equation}
The momentum distribution functions in Eq. (16) with the subscripts suppressed
are expressed as
\begin{equation}\label{distribution}
f(k) =\frac{1+\sum_{l=1}^{\infty} c_l (k\cdot u)^l}{e^{k\cdot u/T}-1},
\end{equation}
where the term $\sum_{l=1}^{\infty} c_l (k\cdot u)^l$ indicates deviation
from equilibrium.

Denote by $\vec{k}^\prime$ the meson momentum in the local reference frame
established on the fluid element, and the meson energy is  
$k^{\prime 0}=k \cdot u$. The number density of
the charmed meson is
\begin{eqnarray}
n=g\int \frac{d^3 k}{(2\pi)^3}f(k)
=\frac{u^0g}{2\pi^2}\int_0^\infty d\mid \vec{k}^\prime \mid
\frac{\vec{k}^{'2}[1+\sum_{l=1}^{\infty}c_l(\sqrt{\vec{k}^{'2}+m^2})^l]}
{e^{\sqrt{\vec{k}^{'2}+m^2}/T}-1} ,
\end{eqnarray}
where $g$ is the degeneracy factor, and $m$ is the mass of the charmed meson.
The thermal-averaged cross section is
\begin{equation}
\langle \sigma_{ij\to i^\prime j^\prime} v_{ij}\rangle
=\frac{\int d^3 k_i^\prime d^3 k_j^\prime
f_i^\prime(k_i^\prime)f_j^\prime(k_j^\prime)\sigma_{ij \to i^\prime j^\prime}
(\sqrt{s}, T)v_{ij}^\prime}
{(u^0)^2\int d^3 k_i^\prime d^3 k_j^\prime f_i^\prime(k_i^\prime)
f_j^\prime(k_j^\prime)},
\end{equation}
with
\begin{equation}
f_i^\prime(k_i^\prime)
=\frac{1+\sum_{l=1}^{\infty} c_l (k_i^{\prime 0})^l}{e^{k_i^{\prime 0}/T}-1},
\end{equation}
\begin{equation}
f_j^\prime(k_j^\prime)
=\frac{1+\sum_{l=1}^{\infty} c_l (k_j^{\prime 0})^l}{e^{k_j^{\prime 0}/T}-1},
\end{equation}
\begin{equation}
v_{ij}^\prime=\frac{\sqrt{(k_i^\prime \cdot k_j^\prime)^2-m_i^2m_j^2}}
{k_i^{\prime0}k_j^{\prime0}}.
\end{equation}

The present work relates to
hadronic matter produced in central nucleus-nucleus collisions. Denote by
($x$, $y$, $z$) the Cartesian coordinates of the fluid element 
(the origin of the local reference frame) in hadronic matter. The four-velocity
of the fluid element takes the form
$u^\mu=\gamma (\frac {t}{\tau},v_r\cos \phi,v_r\sin \phi,\frac {z}{\tau})$ 
where $t$ is the time, $\tau=\sqrt{t^2-z^2}$ the proper time, $v_r$ the 
transverse velocity, and $\gamma =1/\sqrt {1-v^2_r}$ the Lorentz factor. 
The $z$-axis in the rest frame of hadronic matter is set along the moving 
direction of a nucleus, and goes through the nuclear center.
In central collisions hadronic matter has only radial flow and longitudinal
flow. In terms of the proper time and the cylindrical polar coordinates
($r$, $\phi$, $z$), the left-hand side in Eq. (14) becomes
\begin{equation}
\partial_\mu(n_R u^\mu)=\gamma \frac {\partial n_R}{\partial \tau}
+n_R (\frac {\partial \gamma}{\partial \tau} + \frac {\gamma}{\tau})
+\frac {1}{r} \frac {\partial}{\partial r}(rn_R\gamma v_r).
\end{equation}
Hadronic matter possesses
cylindrical symmetry. Equation (24) is then reduced to
\begin{equation}
\partial_\mu(n_R u^\mu)=\gamma \frac {\partial n_R}{\partial \tau}
+\gamma v_r \frac {\partial n_R}{\partial r}
+n_R \gamma^3 v_r \frac {\partial v_r}{\partial \tau}
+n_R \gamma^3 \frac {\partial v_r}{\partial r}
+\frac {n_R \gamma}{\tau} + \frac {n_R\gamma v_r}{r}.
\end{equation}
Combining Eqs. (14), (15), and (25), we get
\begin{align}
&\gamma \frac {\partial n_R}{\partial \tau}
+\gamma v_r \frac {\partial n_R}{\partial r}
+n_R \gamma^3 v_r \frac {\partial v_r}{\partial \tau}
+n_R \gamma^3 \frac {\partial v_r}{\partial r}
+\frac {n_R \gamma}{\tau} + \frac {n_R\gamma v_r}{r}
 \notag \\
&=\langle\sigma_{D\bar{D} \rightarrow \rho R}v_{D\bar D}\rangle n_Dn_{\bar D}
+\langle\sigma_{D\bar{D}^* \rightarrow \pi R}v_{D\bar{D}^*}\rangle n_D
n_{\bar{D}^*}
 \notag \\
&+\langle\sigma_{D^*\bar{D} \rightarrow \pi R}v_{D^*\bar D}\rangle n_{D^*}
n_{\bar D}
+\langle\sigma_{D\bar{D}^* \rightarrow \rho R}v_{D\bar{D}^*}\rangle n_D
n_{\bar{D}^*} \notag \\
&+\langle\sigma_{D^*\bar{D} \rightarrow \rho R}v_{D^*\bar D}\rangle n_{D^*}
n_{\bar D}
+\langle\sigma_{D^*\bar{D}^* \rightarrow \pi R}v_{D^*\bar{D}^*}\rangle n_{D^*}
n_{\bar{D}^*}
\notag \\
&+\langle\sigma_{D^*\bar{D}^* \rightarrow \rho R}v_{D^*\bar{D}^*}\rangle
n_{D^*}n_{\bar{D}^*}. \notag \\
\end{align}

The temperature and the transverse velocity involved in Eq. (26) are
given by the relativistic hydrodynamic equation,
\begin{equation}
\partial_{\mu}T^{\mu\nu}=0,
\end{equation}
where $T^{\mu\nu}$ is the energy-momentum tensor,
\begin{equation}
T^{\mu\nu}=(\epsilon+P)u^{\mu}u^{\nu}-Pg^{\mu\nu}
+\eta [\nabla^\mu u^\nu +\nabla^\nu u^\mu -\frac{2}{3}
(g^{\mu \nu}-u^\mu u^\nu)\nabla \cdot u],
\end{equation}
where $\epsilon$ is the energy density, $P$ the pressure, $g^{\mu \nu}$ the 
metric, $\eta$ the shear viscosity, 
and $\nabla^\mu =\partial^\mu -u^\mu u \cdot \partial$. A
parametrization of the shear viscosity is given in Ref. \cite{NEP}.

\vspace{0.5cm}
\centerline{\bf IV. NUMERICAL RESULTS AND DISCUSSIONS }
\vspace{0.5cm}

Hadronic matter created in ultrarelativistic heavy-ion collisions changes 
during expansion. Numerical cross sections obtained from Eq. (13) show 
remarkable
temperature dependence. $\psi(4040)$, $\psi(4160)$, and $\psi(4415)$ mesons
produced in hadronic matter are appropriate to measurements.

\centerline{\bf A. Cross sections}

The potential given in Eq. (7) at large distances decreases with increasing 
temperature, i.e., confinement becomes weaker and weaker.
We solve the Schr\"odinger equation with the potential to obtain meson masses
and mesonic quark-antiquark relative-motion wave functions in coordinate 
space. The wave functions depend on temperature, and give meson radii which 
increase with increasing temperature. From the Fourier transform we get
the potential and the wave functions in momentum space, which are used in
Eqs. (4) and (6).
We calculate transition amplitudes in the prior form and in the post
form, which lead to unpolarized cross sections. The unpolarized cross sections
depend on temperature, and are plotted in Figs. 4-18 for the fifteen reactions:
\begin{displaymath}
D \bar{D} \to \rho \psi(4040),~D \bar{D} \to \rho \psi(4160),
~D \bar{D} \to \rho \psi(4415),
\end{displaymath}
\begin{displaymath}
D \bar{D}^* \to \pi \psi(4040),~D \bar{D}^* \to \pi \psi(4160),
~D \bar{D}^* \to \pi \psi(4415),
\end{displaymath}
\begin{displaymath}
D \bar{D}^* \to \rho \psi(4040),~D \bar{D}^* \to \rho \psi(4160),
~D \bar{D}^* \to \rho \psi(4415),
\end{displaymath}
\begin{displaymath}
D^* \bar{D}^* \to \pi \psi(4040),~D^* \bar{D}^* \to \pi \psi(4160),
~D^* \bar{D}^* \to \pi \psi(4415),
\end{displaymath}
\begin{displaymath}
D^* \bar{D}^* \to \rho \psi(4040),~D^* \bar{D}^* \to \rho \psi(4160),
~D^* \bar{D}^* \to \rho \psi(4415).
\end{displaymath}
The cross sections for $D^*\bar D$ reactions are identical to those for 
$D\bar{D}^*$ reactions. Since the isospin quantum numbers of $\psi(4040)$, 
$\psi(4160)$,
and $\psi(4415)$ are zero, the total isospin of final mesons is that of the
pion or the $\rho$ meson.

The reactions that have $\rho$ mesons in the 
final states are endothermic reactions. The cross section for every endothermic
reaction at a given temperature has a maximum, i.e., every cross-section
curve has at least one peak. Weakening confinement leads to decreasing 
cross
section with increasing temperature. However, increase of initial-meson
radii cause increase of cross sections. The two factors generate variation 
of the peak cross section with increasing temperature. At zero temperature the
peak cross sections of $D \bar{D} \rightarrow \rho \psi(4160)$ 
($D \bar{D}^* \rightarrow \rho \psi(4160)$, 
$D^* \bar{D}^* \rightarrow \rho \psi(4160)$)
and $D \bar{D} \rightarrow \rho \psi(4415)$
($D \bar{D}^* \rightarrow \rho \psi(4415)$,
$D^* \bar{D}^* \rightarrow \rho \psi(4415)$) are similar, but smaller than that
of $D \bar{D} \rightarrow \rho \psi(4040)$
($D \bar{D}^* \rightarrow \rho \psi(4040)$,
$D^* \bar{D}^* \rightarrow \rho \psi(4040)$). At any of the five temperatures
$T/T_{\rm c}=0.65$, 0.75, 0.85, 0.9, or 0.95, the peak cross section of
$D \bar{D} \rightarrow \rho \psi(4040)$
($D \bar{D}^* \rightarrow \rho \psi(4040)$,
$D^* \bar{D}^* \rightarrow \rho \psi(4040)$) is similar to or larger than
the one of
$D \bar{D} \rightarrow \rho \psi(4415)$ 
($D \bar{D}^* \rightarrow \rho \psi(4415)$, 
$D^* \bar{D}^* \rightarrow \rho \psi(4415)$), and the one of
$D \bar{D} \rightarrow \rho \psi(4160)$ 
($D \bar{D}^* \rightarrow \rho \psi(4160)$, 
$D^* \bar{D}^* \rightarrow \rho \psi(4160)$) is generally much smaller than 
the latter.

The reactions that produce pions are endothermic at $T=0$ and exothermic at
$T/T_{\rm c}=0.65$, 0.75, 0.85, 0.9, and 0.95. For the endothermic reactions,
the peak cross section of $D \bar{D}^* \rightarrow \pi \psi(4040)$ is larger
than the one of $D \bar{D}^* \rightarrow \pi \psi(4160)$ which is larger than
that of $D \bar{D}^* \rightarrow \pi \psi(4415)$. Away from threshold 
energies, every cross-section curve of the exothermic reactions has one or two
maxima. The peak cross section of producing $\psi(4040)$ is largest and
the one of producing $\psi(4160)$ is smallest among the reactions
$D \bar{D}^* \rightarrow \pi \psi(4040)$,
$D \bar{D}^* \rightarrow \pi \psi(4160)$, and
$D \bar{D}^* \rightarrow \pi \psi(4415)$. A similar case holds true for
$D^* \bar{D}^* \rightarrow \pi \psi(4040)$,
$D^* \bar{D}^* \rightarrow \pi \psi(4160)$, and
$D^* \bar{D}^* \rightarrow \pi \psi(4415)$.

Numerical cross sections plotted in Figs. 4-18 are used in the master rate
equations. For convenience they are parametrized as
\begin{eqnarray}
\sigma^{\rm unpol}(\sqrt {s},T) & = &
a_1 \left( \frac {\sqrt {s} -\sqrt {s_0}}{b_1} \right)^{c_1}
\exp \left[ c_1 \left( 1-\frac {\sqrt {s} -\sqrt {s_0}}{b_1} \right) \right]
                   \notag   \\
& & + a_2 \left( \frac {\sqrt {s} -\sqrt {s_0}}{b_2} \right)^{c_2}
\exp \left[ c_2 \left( 1-\frac {\sqrt {s} -\sqrt {s_0}}{b_2} \right) \right] ,
\end{eqnarray}
for endothermic reactions and
\begin{eqnarray}
\sigma^{\rm unpol}(\sqrt {s},T) & = &
\frac{{\vec {P'}}^2}{{\vec P}^2}
\left\{a_1 \left( \frac {\sqrt {s} -\sqrt {s_0}}{b_1} \right)^{c_1}
\exp \left[ c_1 \left( 1-\frac {\sqrt {s} -\sqrt {s_0}}{b_1}
\right) \right]\right.
                   \notag   \\
& & \left. + a_2 \left( \frac {\sqrt {s} -\sqrt {s_0}}{b_2} \right)^{c_2}
\exp \left[ c_2 \left( 1-\frac {\sqrt {s} -\sqrt {s_0}}{b_2}
\right) \right]\right\} ,
\end{eqnarray}
for exothermic reactions. $\sqrt{s}_0$ is the threshold energy, which equals 
the sum of the masses of
the two initial (final) mesons for the exothermic (endothermic) reaction.
The temperature-dependent charmonium masses obtained from the Schr\"odinger
equation are parametrized as
\begin{equation}
m_{\psi (4040)}=3.59 \left[ 1-\left( \frac{T}{1.26T_{\rm c}} \right)^{3.16}
\right]^{0.35},
\end{equation}
\begin{equation}
m_{\psi (4160)}=3.64 \left[ 1-\left( \frac{T}{1.34T_{\rm c}} \right)^{3.52}
\right]^{0.54},
\end{equation}
\begin{equation}
m_{\psi (4415)}=3.74 \left[ 1-\left( \frac{T}{1.21T_{\rm c}} \right)^{2.36}
\right]^{0.25}.
\end{equation}

In hadronic matter where the temperature is constrained by $0.6T_{\rm c} \leq
T<T_{\rm c}$, the masses of $\psi(4040)$, $\psi(4160)$, and $\psi(4415)$
are smaller than the sum of two open-charm mesons. Therefore, 
the production of the three charmonia from fusion of two open-charm mesons
is not allowed \cite{LXW}.

The transition amplitudes for scattering in the prior form and in the post
form include the Fourier transform of the coordinate-space potential given 
in Eq. (7):
\begin{eqnarray}
V_{ab}\left( \vec {Q}\right) &=& -\frac{ \vec {\lambda }_{a}}{2}
\cdot \frac{\vec {\lambda }_{b}}{2}\xi_1
\left[ 1.3- \left( \frac {T}{T_{\rm c}} \right)^4 \right]
\left[ (2\pi)^3\delta^3 (\vec {Q}) - \frac {8\pi}{Q}
\int^\infty_0 dr \frac {r\sin (Qr)}{\exp (2\xi_2 r)+1} \right]
                               \notag \\
& &
+\frac{ \vec {\lambda }_{a}}{2} \cdot \frac{\vec {\lambda }_{b}}{2} 64 \pi\xi_3
\int^\infty_0 dq \frac {\rho (q^2) 
-\frac {K}{q^2}}{(\xi_3^2+Q^2+q^2)^2-4Q^2q^2}
                               \notag \\
& & -\frac{\vec {\lambda }_{a}}{2}
\cdot \frac{\vec {\lambda }_{b}}{2}\frac{16\pi ^{2}}{25} 
\exp \left( -\frac {Q^2}{4d^2} \right)
\frac{\vec {s}_{a}\cdot \vec {s}_{b}}{m_{a}m_{b}}
                               \notag \\
& &
+\frac{\vec {\lambda }_{a}}{2}\cdot \frac{\vec {\lambda }_{b}}{2}
\frac{16\pi ^{2}\lambda }{25Q}\int_{0}^{\infty}dx\frac{d^2v(x) }{dx^2}
\sin \left( \frac{Q}{\lambda }x\right)
\frac{\vec {s}_{a}\cdot \vec {s}_{b}}{m_{a}m_{b}} 
                                 \notag \\
& &
-\frac{\vec {\lambda }_{a}}{2}\cdot \frac{\vec {\lambda }_{b}}{2}
\frac{24\pi ^{2}\lambda }{25Q}\int_{0}^{\infty}dx \frac{1}{x^2}
\left[ v(x)-x\frac{dv(x)}{dx}+\frac{x^2}{3}\frac{d^2v(x) }{dx^2} \right]
                                 \notag \\
& &
\left[ \left( 1-\frac{3\lambda^2}{Q^2x^2} \right)
\sin \left( \frac{Q}{\lambda }x\right)
+\frac{3\lambda}{Qx} \cos \left( \frac{Q}{\lambda }x\right) \right]
\frac{3s_{az}s_{bz}-\vec {s}_{a}\cdot \vec {s}_{b}}{m_{a}m_{b}} ,
                                 \notag \\
\end{eqnarray}
where $\vec Q$ is the momentum attached to the dot-dashed lines in Figs. 1 and 
2, $K=3/16\pi^2\alpha^\prime$, $\rho (q^2)$ is the physical running coupling 
constant \cite{BT}, and $s_{az}$ ($s_{bz}$) is the magnetic projection
quantum number of $\vec{s}_a$ ($\vec{s}_b$). The expressions of
$\cal M_{\rm fi}^{\rm prior}$ and $\cal M_{\rm fi}^{\rm post}$ with this
momentum-space potential involve
seven-dimensional integrals. Because of the two sines and the cosine in the 
fourth and fifth terms,
in a short computational time to carry out integration at a given temperature
and a given value of $\sqrt s$ is impossible. In order to reduce the
computational
time, the two integrals in the fourth and fifth terms are parametrized as
\begin{equation}
\int_{0}^{\infty}dx\frac{d^2v(x) }{dx^2}\sin \left( \frac{Q}{\lambda }x\right)
=(1-0.36e^{-0.1Q}-0.58e^{-0.4Q}-0.06e^{-0.4Q^2})(-1.7-0.0575Q) ,
\end{equation}
\begin{eqnarray}
& & \int_{0}^{\infty}dx \frac{1}{x^2}
\left[ v(x)-x\frac{dv(x)}{dx}+\frac{x^2}{3}\frac{d^2v(x) }{dx^2} \right]
\left[ \left( 1-\frac{3\lambda^2}{Q^2x^2} \right)
\sin \left( \frac{Q}{\lambda }x\right)
+\frac{3\lambda}{Qx} \cos \left( \frac{Q}{\lambda }x\right) \right]
           \nonumber \\
& &
= (1-0.59e^{-0.15Q}-0.31e^{-0.3Q^2}-0.1e^{-6.71Q^2})(-0.568-0.0375Q),
\end{eqnarray}
where the unit of $Q$ is ${\rm fm}^{-1}$. The momentum-space potential with 
the two
parametrizations causes the post-prior discrepancy. To know the contribution
from scattering in the prior form or in the post form to the unpolarized
cross section $\sigma^{\rm unpol}$, in Fig. 19 we compare the cross sections
($\sigma_{\rm unpol}^{\rm prior}$ and $\sigma_{\rm unpol}^{\rm post}$) for
scattering in the prior form and in the post form with the unpolarized
cross section ($\sigma^{\rm unpol}$) for the
exothermic reaction $D\bar{D}^* \to \pi\psi(4040)$ at $T/T_{\rm c}=0.85$ and 
the endothermic reaction $D\bar{D}^* \to \rho\psi(4040)$. At few 
energies, for example, $\sqrt{s}=3.765$ GeV in the right panel, the dotted 
curves and the dashed curves cross, i.e., 
$\sigma_{\rm unpol}^{\rm prior}=\sigma_{\rm unpol}^{\rm post}
=\sigma^{\rm unpol}$, which indicates that scattering
in the prior form and in the post form make the same contribution to
$\sigma^{\rm unpol}$. When $\sqrt s$ increases from
3.6 GeV, the dotted curves approach the dashed curves, which means that the 
post-prior discrepancy becomes smaller and smaller. 
The threshold energy of $D\bar{D}^* \to \pi\psi(4040)$ 
at $T/T_{\rm c}=0.85$ is 3.45709 GeV. At $\sqrt {s}=3.4571$ GeV
the cross section for scattering in the prior form is by 81.7\% smaller than 
$\sigma^{\rm unpol}$, while the cross section for 
scattering in the post form is by 81.7\% larger than $\sigma^{\rm unpol}$. 
At $\sqrt {s}=3.47209$ GeV the 
solid curve exhibits a maximum, and 
$\sigma_{\rm unpol}^{\rm prior}$ ($\sigma_{\rm unpol}^{\rm post}$) is by 68.7\%
larger (smaller) than $\sigma^{\rm unpol}$. In the right panel, 
$\sigma^{\rm unpol}$ for
$D\bar{D}^* \to \rho\psi(4040)$ at $\sqrt {s}=3.51319$ GeV (3.52569 GeV,
3.55569 GeV) has a maximum (minimum, maximum), and 
$\sigma_{\rm unpol}^{\rm prior}$ and $\sigma_{\rm unpol}^{\rm post}$ deviate
from $\sigma^{\rm unpol}$ by 5.6\% (87\%, 59.7\%). Contributions from 
scattering 
in the prior form and in the post form change with increasing $\sqrt s$.

The potential used by Barnes and Swanson in Ref. \cite{BS} includes the color 
Coulomb interaction, the linear confining interaction, and the 
one-gluon-exchange
spin-spin hyperfine interaction. Describing quark-antiquark relative motion
in all mesons by a Gaussian wave function, they obtained that the cross section
for elastic $\pi \pi$ ($KK$) scattering for total isospin $I=2$ ($I=1$) in the 
post form equals the
one in the prior form. This means that the scattering amplitude in the post
form equals the scattering amplitude in the prior form even though the 
Gaussian wave function is not the exact quark-antiquark relative-motion
wave functions that are solutions of the Schr\"odinger equation with the 
potential. Matrix elements of the spin-spin hyperfine interaction in
scattering in the prior 
form are given by Eqs. (71)-(74) in Ref. \cite{BS} when the four constituents
of scattering mesons have the same mass. However,
the four equations are not enough to help us in understanding that the 
scattering amplitude in the post form equals the one
in the prior form. Therefore, in Ref. \cite{LXL} we present analytic
expressions of all matrix elements of the spin-spin interaction in 
scattering in the post form and in the prior form allowing that the four
constituents have different masses and that different Gaussian wave functions 
are used to describe quark-antiquark relative motion of different mesons.
Every matrix element is the product of a flavor matrix element, 
a color matrix element, a spin matrix element,
and a spatial matrix element. The two upper diagrams
in Fig. 1 (Fig. 2) are capture diagrams in which the interacting
quark-antiquark pair scatter into (come from) the same final (initial) meson.
The four lower diagrams in Figs. 1 and 2 are transfer diagrams in which
the two interacting quarks or antiquarks scatter into different final
mesons. The transfer diagrams in the post and prior
forms are identical, and the capture diagrams in the two forms look different.
A matrix element is associated with a diagram. When the sum of the two matrix
elements corresponding to the capture diagrams in the post form does not
equal the sum in the prior form, the post-prior discrepancy occurs.
The same flavor matrix element is applied to the eight diagrams in Figs. 1 and 
2. The color 
matrix elements associated with the four capture diagrams are all -4/9.
For elastic scattering of two pseudoscalar mesons, the spin 
matrix elements corresponding to the four capture diagrams are the same. Hence,
the matrix element of the spin-spin interaction in a (the other)
capture diagram in the post form equals the one in a (the other)
capture diagram in the prior form, if the Gaussian wave functions of the 
initial and final mesons are identical and the quarks or the antiquarks of
scattering mesons have the same flavor \cite{LXL}.
Numerical calculations show that the matrix element of the central 
spin-independent potential in the post form equals the one in the prior 
form \cite{LXL}. Therefore, the scattering amplitude in the post form equals 
the scattering amplitude in the prior form, and no post-prior discrepancy
happens in elastic scattering of two pseudoscalar mesons \cite{BS,LXL},
when the quark-antiquark relative motion of the initial and final mesons are
described by the same Gaussian wave function and if the 
quarks or the antiquarks of scattering mesons have the same flavor.
No post-prior discrepancy is also true for elastic scattering between two 
vector mesons or between a pseudoscalar meson and a vector meson \cite{LXL}.

In Ref. \cite{BS} $H_0(A)$, $H_0(B)$, $H_0(C)$, and $H_0(D)$ of mesons $A$,
$B$, $C$, and $D$ in the scattering $A+B \to C+D$ individually consist of
the kinetic energies of the quark and the antiquark and the potential between
the quark and the antiquark. The Schr\"odinger equation with $H_0$ provides
masses and exact quark-antiquark relative-motion wave functions of 
the four mesons. The interaction $H_I(A,B)$ between the two 
constituents of meson $A$ and the two constituents of meson $B$ turns the
color-singlet states $A$ and $B$ into color-octet states. During propagation
of quarks and antiquarks, quark interchange causes a quark and an antiquark to
form the color-singlet state $C$ as well as the other quark and the other 
antiquark to form the color-singlet state $D$. This decomposition of 
the Hamiltonian of the four constituents, 
$H=H_0(A)+H_0(B)+H_I(A,B)$, reflects the prior form of scattering. Certainly,
quark interchange between mesons $A$ and $B$ can take place before the 
interaction
$H_I(C,D)$ between the two constituents of meson $C$ and the two constituents
of meson $D$, and breaks up mesons $A$ and $B$ 
to yield two color-octet states. The interaction $H_I(C,D)$ makes the two
color-octet states colorless so that the bound states $C$ and $D$ are formed.
Scattering in the post form thus gives rise to this decomposition 
$H=H_0(C)+H_0(D)+H_I(C,D)$.
Therefore, the post-prior discrepancy (originally the $H_I(A,B)$ matrix element
does not equal the $H_I(C,D)$ matrix element) is related to the decomposition
of the Hamiltonian. 

If two or more mesons in elastic meson-meson scattering are described by
different Gaussian wave functions, the scattering amplitudes in the post
form and in the prior form are not the same, and the post-prior discrepancy
takes place \cite{LXL}. 
For inelastic scattering of two mesons, the spin matrix elements associated
with the four capture diagrams can not guarantee that the sum of the two matrix
elements corresponding to the capture diagrams in the post form 
equals the sum in the prior form, the scattering amplitudes in the two
forms are thus not identical,
and the post-prior discrepancy occurs \cite{BS}. 
If the exact mesonic quark-antiquark
wave functions are used, no post-prior discrepancy exists. However, the exact
wave functions are not amenable to analytic calculations. To get analytic
expressions of spatial matrix elements, the sum of several Gaussian wave 
functions was suggested in Ref. \cite{BS} to approach the exact quark-antiquark
relative-motion wave functions. The more Gaussian wave functions
that are used, the smaller post-prior discrepancy that is observed.

\centerline{\bf B. Number densities}

Since charmed mesons are well measured in Pb-Pb collisions at the LHC, we get
their distribution functions $f(k)$ from experimental data. The
Cooper-Frye formula \cite{CF} is
\begin{equation}
E\frac{d^3N}{d^3 k}=\frac{g}{(2\pi)^3}\int_{\sigma_f} f(k)k^\mu d\sigma_\mu ,
\end{equation}
where $E$, $k^\mu$, and $N$ are the energy, the four-momentum, and the number
of the charmed meson, respectively;
$\sigma_f$ is the freeze-out surface with the normal vector
$d\sigma_\mu$. With the space-time rapidity $\eta = \frac{1}{2}\ln
\frac{t+z}{t-z}$, we get
\begin{eqnarray}
\frac{dN}{dk_T}&=&\frac{g}{(2\pi)^3}\int_{R_{\rm c}}^{R_{\rm fz}}\int_0^{2\pi}
\int_{-10}^{10}\int_{0}^{2\pi}\int_{y_{\rm min}}^{y_{\rm max}} f(k)
[m_T\cosh (y-\eta)-k_Tv_r \cos (\phi -\varphi )]  \nonumber \\
& & k_T\tau r drd\phi d\eta d\varphi dy ,
\end{eqnarray}
where $y$, $\vec{k}_T$, and $m_T$ are the rapidity, the transverse momentum,
and the transverse mass of the charmed meson, respectively; $\varphi$ is 
the angle between the transverse momentum and the $x$-axis; $R_{\rm c}$ and
$R_{\rm fz}$ are the radii of hadronic-matter surface at hadronization and at 
kinetic freeze-out, respectively.
In Ref. \cite{ALICE5020D} the $D$-meson production yields are measured at
midrapidity ($\mid y \mid <0.5$) as functions of transverse momentum.
$y_{\rm min}$ and $y_{\rm max}$ are thus set as -0.5 and 0.5, respectively.
Fits to the experimental data of $dN/dp_T$ of prompt $D^+$, $D^0$, and
$D^{*+}$ mesons at $p_T<8$ GeV/$c$ in central Pb-Pb collisions at 
$\sqrt{s_{NN}}=5.02$ TeV with $T=0.1686$ GeV give
\begin{displaymath}
l=11,~~~~~c_l=6\times 10^{-12},~~~~~{\rm for}~ D^+~ {\rm meson};
\end{displaymath}
\begin{displaymath}
l=9,~~~~~c_l=3\times 10^{-9},~~~~~{\rm for}~D^0~ {\rm meson};
\end{displaymath}
\begin{displaymath}
l=14,~~~~~c_l=5\times 10^{-16},~~~~~{\rm for} ~D^{*+}~ {\rm meson};
\end{displaymath}
$c_l$ equal zero for other $l$ values.
This means that only one term in the sum
$\sum_{l=1}^{\infty} c_l (k \cdot u)^l$ is needed for each $D$ meson.

Quark-gluon matter initially produced in Pb-Pb collisions at LHC energies is 
not a thermal state. Undergoing elastic parton-parton-parton scattering and 
parton-parton scattering, quark-gluon matter acquires a temperature in a short 
time and becomes a quark-gluon plasma\cite{XuWWND,XuHEP,Xu2013}. Hydrodynamic 
models are applied to the quark-gluon plasma\cite{Florkowski,Ryblewski}. For 
central Pb-Pb collisions at 
$\sqrt{s_{NN}}$=5.02~\rm{TeV} we get $0.82~\rm{GeV}$ as the initial 
temperature of the quark-gluon plasma at a time of the
order of 0.65 fm/$c$ and 10.05 fm/$c$ as the proper time at which 
hadronization of the quark-gluon plasma occurs. We start solving
the hydrodynamic equation (Eq. (27)) for hadronic matter
at the time 10.05 fm/$c$ with $T_c=0.175$ GeV.
Since $\psi(4040)$, $\psi(4160)$, and $\psi(4415)$ mesons are dissolved in 
hadronic 
matter when the temperature is larger than $0.97T_{\rm c}$, $0.95T_{\rm c}$ 
and $0.87T_{\rm c}$, respectively, their number densities are zero above the
three dissociation temperatures. From the dissociation temperatures,
Eq. (26) is solved until
kinetic freeze-out to get number densities that are functions of the proper
time and the radius. Variation of the number densities with respect to the
proper time at $r=0$ fm is
drawn in Fig. 20, and radius dependence at kinetic freeze-out is plotted
in Fig. 21. Hadronic matter produced from the quark-gluon plasma expands, and
its temperature decreases from the critical temperature. When the temperature
arrives at 0.97$T_{\rm c}$, 0.95$T_{\rm c}$, and 0.87$T_{\rm c}$, production of
$\psi(4040)$, $\psi(4160)$, and $\psi(4415)$ get started from reactions
of charmed mesons, respectively. Therefore, when the proper time increases 
from 10.83 fm/$c$, 11.38 fm/$c$, and 13.95 fm/$c$, respectively,
the number densities of $\psi(4040)$, $\psi(4160)$, and $\psi(4415)$ 
increase. However, the three mesons produced at $r=0$ fm spread out, and this
reduces the number densities. When the reduced amount exceeds the increased
amount, the number densities decrease as seen in Fig. 20. In Fig. 21 the
number densities decrease slowly when $r$ increases from zero. 
From the last subsection, we have already known that at $T=0.65T_{\rm c}$,
$0.75T_{\rm c}$, $0.85T_{\rm c}$, $0.9T_{\rm c}$, and $0.95T_{\rm c}$
the peak cross sections of $D\bar{D} \to \rho \psi(4040)$
($D\bar{D}^* \to \pi \psi(4040)$, $D\bar{D}^* \to \rho \psi(4040)$,
$D^*\bar{D}^* \to \pi \psi(4040)$, $D^*\bar{D}^* \to \rho \psi(4040)$) are
similar to or larger than the ones of $D\bar{D} \to \rho \psi(4415)$
($D\bar{D}^* \to \pi \psi(4415)$, $D\bar{D}^* \to \rho \psi(4415)$,
$D^*\bar{D}^* \to \pi \psi(4415)$, $D^*\bar{D}^* \to \rho \psi(4415)$), and 
the latter are generally larger than those of $D\bar{D} \to \rho \psi(4160)$
($D\bar{D}^* \to \pi \psi(4160)$, $D\bar{D}^* \to \rho \psi(4160)$,
$D^*\bar{D}^* \to \pi \psi(4160)$, $D^*\bar{D}^* \to \rho \psi(4160)$). 
Therefore, the number density of $\psi(4040)$ is larger than the one of 
$\psi(4415)$ which is larger than that of $\psi(4160)$. At kinetic freeze-out,
hadronic matter has a volume of the order of $6\times 10^4$ ${\rm fm}^3$. 
As a consequence of the number densities, 
the numbers of $\psi(4040)$, $\psi(4160)$, and $\psi(4415)$ at kinetic 
freeze-out are 0.25, 0.1, and 0.18 , respectively.

\vspace{0.5cm}
\leftline{\bf V. SUMMARY }
\vspace{0.5cm}

We have studied the production of $\psi(4040)$, $\psi(4160)$, and $\psi(4415)$
mesons in collisions of charmed mesons in the quark interchange mechanism and
space-time distribution of the three charmonia in hadronic matter created in 
ultrarelativistic heavy-ion collisions with the master rate equations.
Formulas of the transition amplitudes are given explicitly. The temperature
dependence of the quark potential, the meson masses, and the mesonic 
quark-antiquark relative-motion wave functions lead to remarkable temperature
dependence of unpolarized cross sections. We have obtained the unpolarized
cross sections for $D\bar{D} \to \rho R$, $D\bar{D}^* \to \pi R$, 
$D\bar{D}^* \to \rho R$, $D^*\bar{D} \to \pi R$, $D^*\bar{D} \to \rho R$,
$D^*\bar{D}^* \to \pi R$, and $D^*\bar{D}^* \to \rho R$, where $R$ denotes
$\psi(4040)$, $\psi(4160)$, or $\psi(4415)$. In a collision of two charmed
mesons, the peak cross sections of producing $\psi(4040)$
at $T=0.65T_{\rm c}$, $0.75T_{\rm c}$, $0.85T_{\rm c}$, $0.9T_{\rm c}$, 
and $0.95T_{\rm c}$ are similar
to or larger than the ones of producing $\psi(4415)$, and the latter are 
generally larger than those of producing $\psi(4160)$. The numerical cross
sections are parametrized so that they can be conveniently used in the master
rate equations. For hadronic matter created in central nucleus-nucleus
collisions, the master rate equations are given in terms of the proper time and
the cylindrical polar coordinates. Below the dissociation temperatures of
$\psi(4040)$, $\psi(4160)$, and $\psi(4415)$ the twenty-one reactions
produce the three charmonia, 
and the master rate equations give number densities
that first increase with increasing time. Among the 
number densities of $\psi(4040)$, $\psi(4160)$, and $\psi(4415)$, at kinetic
freeze-out the one of $\psi(4040)$ is largest and that of $\psi(4160)$ is
smallest. One can find the three charmonia in the large volume of hadronic 
matter created in central Pb-Pb collisions at $\sqrt{s_{NN}}=5.02$ TeV.

\vspace{0.5cm}
\leftline{\bf APPENDIX A: DERIVE THE ISOSPIN-AVERAGED CROSS SECTION}
\vspace{0.5cm}

Denote the isospin of meson $A$ ($B$, $C$, $D$) by $I_{A}$ ($I_{B}$, $I_{C}$,
$I_{D}$) and its $z$ component by $I_{Az}$ ($I_{Bz}$, $I_{Cz}$, $I_{Dz}$). 
Let $\phi_{A\rm{flavor}}^{I_{A}I_{Az}}$, $\phi_{B\rm{flavor}}^{I_{B}I_{Bz}}$,
$\phi_{C\rm{flavor}}^{I_{C}I_{Cz}}$, and $\phi_{D\rm{flavor}}^{I_{D}I_{Dz}}$
be the flavor wave functions of mesons $A$, $B$, $C$, and $D$, respectively.
$\phi_{A\rm{flavor}}^{I_{A}I_{Az}}$ and
$\phi_{B\rm{flavor}}^{I_{B}I_{Bz}}$
are coupled to the flavor wave function $\varphi_{AB\rm{flavor}}^{II_{z}}$
with the total isospin $I$ and its $z$ component $I_{z}$.
$\phi_{C\rm{flavor}}^{I_{C}I_{Cz}}$ and $\phi_{D\rm{flavor}}^{I_{D}I_{Dz}}$
are coupled to the flavor wave function
$\varphi_{CD\rm{flavor}}^{I^{\prime}I_{z}^{\prime}}$
with the total isospin $I^{\prime}$ and its $z$ component $I_{z}^{\prime}$:
\begin{eqnarray}
\phi_{A\rm{flavor}}^{I_{A}I_{Az}}\phi_{B\rm{flavor}}^{I_{B}I_{Bz}}
=\sum_{II_{z}}(I_{A}I_{Az}I_{B}I_{Bz} \mid II_{z})
\varphi_{AB\rm{flavor}}^{II_{z}},
\end{eqnarray}
\begin{eqnarray}
\phi_{C\rm{flavor}}^{I_{C}I_{Cz}}\phi_{D\rm{flavor}}^{I_{D}I_{Dz}}
=\sum_{I^{\prime}I_{z}^{\prime}}(I_CI_{Cz}I_DI_{Dz} \mid 
I^{\prime}I_{z}^{\prime})
\varphi_{CD\rm{flavor}}^{I^{\prime}I_{z}^{\prime}},
\end{eqnarray}
where $(I_{A}I_{Az}I_{B}I_{Bz} \mid II_{z})$ and
$(I_{C}I_{Cz}I_{D}I_{Dz} \mid I^{\prime}I_{z}^{\prime})$ are the Clebsch-Gordan
coefficients.
The isospin part of the transition amplitudes is 
\begin{equation}
{\cal{M}_{\rm{isospin}}}  
=[\phi_{C\rm{flavor}}^{I_{C}I_{Cz}}\phi_{D\rm{flavor}}^{I_{D}I_{Dz}}]^{+}
P_{\rm qi}
\phi_{A\rm{flavor}}^{I_{A}I_{Az}}\phi_{B\rm{flavor}}^{I_{B}I_{Bz}},
\end{equation}
where the symbol $P_{\rm qi}$ is the operator that implements quark interchange
in flavor space. 
\begin{eqnarray}
\sum_{I_{Az}I_{Bz}I_{Cz}I_{Dz}}|{\cal{M}_{\rm{isospin}}}|^{2}  
&=&\sum_{I_{Az}I_{Bz}I_{Cz}I_{Dz}}|
[\phi_{C\rm{flavor}}^{I_{C}I_{Cz}}\phi_{D\rm{flavor}}^{I_{D}I_{Dz}}]^{+}
P_{\rm qi}
\phi_{A\rm{flavor}}^{I_{A}I_{Az}}\phi_{B\rm{flavor}}^{I_{B}I_{Bz}}|^{2}
\nonumber\\
&=&\sum_{I^{\prime}I_{z}^{\prime}II_{z}}|
\varphi_{CD\rm{flavor}}^{I^{\prime}I_{z}^{\prime}+}P_{\rm qi}
\varphi_{AB\rm{flavor}}^{II_{z}}|^{2}.
\end{eqnarray}
Because of isospin conservation, $I^{\prime}=I$ and $I_{z}^{\prime}=I_{z}$,
\begin{eqnarray}
\sum_{I_{Az}I_{Bz}I_{Cz}I_{Dz}}|{\cal{M}_{\rm{isospin}}}|^{2}
=\sum_{II_{z}}|\varphi_{CD\rm{flavor}}^{II_{z}+}P_{\rm qi}
\varphi_{AB\rm{flavor}}^{II_{z}}|^{2}.
\end{eqnarray}
Since flavor matrix elements are independent of $I_{z}$,
\begin{eqnarray}
\sum_{I_{Az}I_{Bz}I_{Cz}I_{Dz}}|{\cal{M}_{\rm{isospin}}}|^{2}
=\sum_{I}(2I+1)|\varphi_{CD\rm{flavor}}^{II+}P_{\rm qi}
\varphi_{AB\rm{flavor}}^{II}|^{2}.
\end{eqnarray}
Finally, the average over the isospin states of the two initial mesons and the
sum over the isospin states of the two final mesons lead to
the isospin-averaged unpolarized cross section for $A+B\rightarrow C+D$,
\begin{eqnarray}
\sigma^{\rm{isoav}}(\sqrt{s},T)=\frac{1}{(2I_{A}+1)(2I_{B}+1)}
\sum_{I}(2I+1)\sigma^{\rm{unpol}}(I, \sqrt{s},T),
\end{eqnarray}
where $\sigma^{\rm unpol}(I,\sqrt{s},T)$ is given in Eq. (13) for the $I$ 
channel.

\vspace{0.5cm}
\leftline{\bf ACKNOWLEDGEMENTS}
\vspace{0.5cm}

This work was supported by the project STRONG2020 of European Center for
Theoretical Studies in Nuclear Physics and Related Areas.

\newpage
\begin{figure}[htbp]
  \centering
    \includegraphics[scale=0.8]{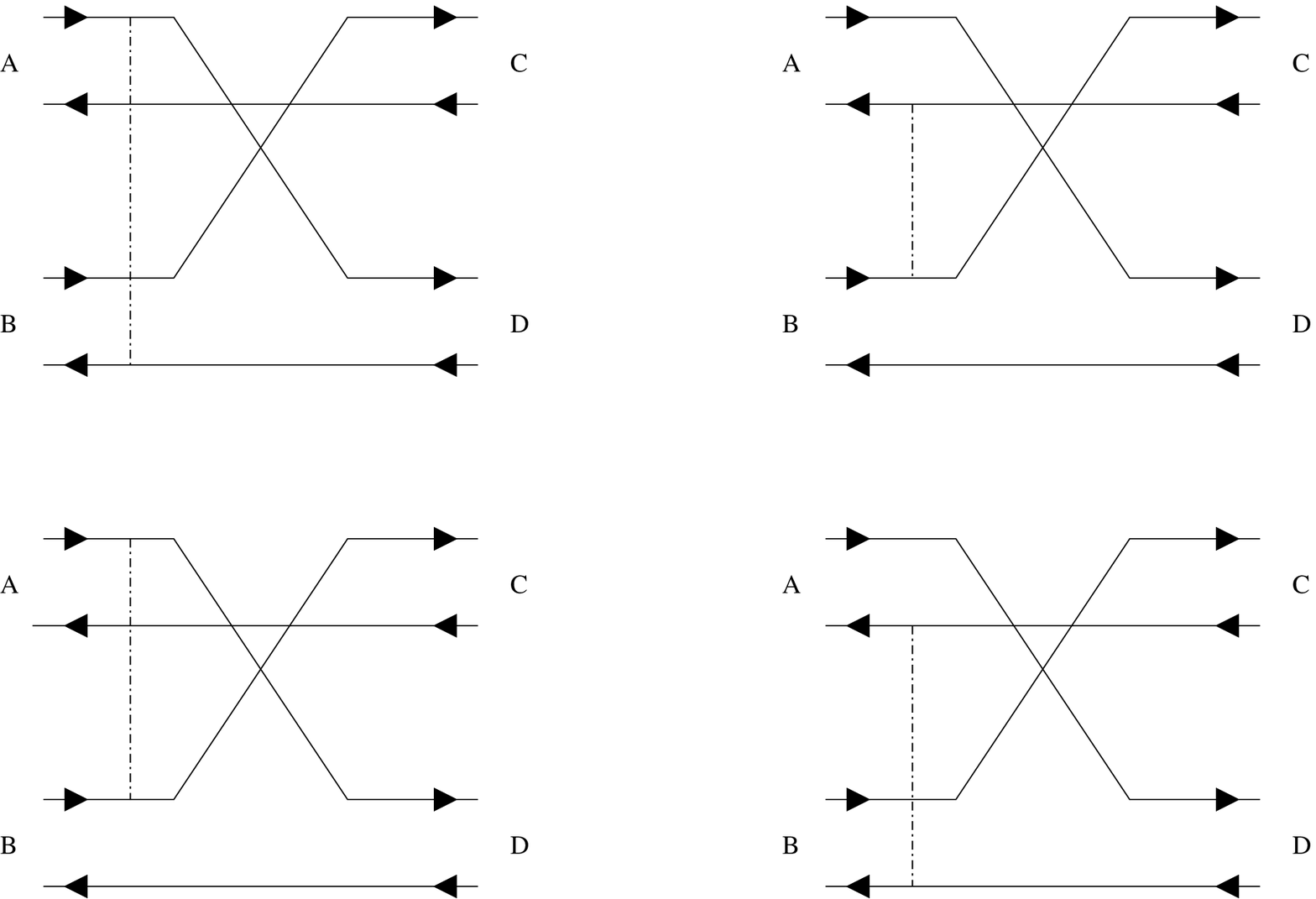}
\caption{Scattering in the prior form. Solid lines with triangles right (left) 
represent quarks (antiquarks). Dot-dashed lines indicate
interactions.}
\label{fig1}
\end{figure}

\newpage
\begin{figure}[htbp]
  \centering
    \includegraphics[scale=0.8]{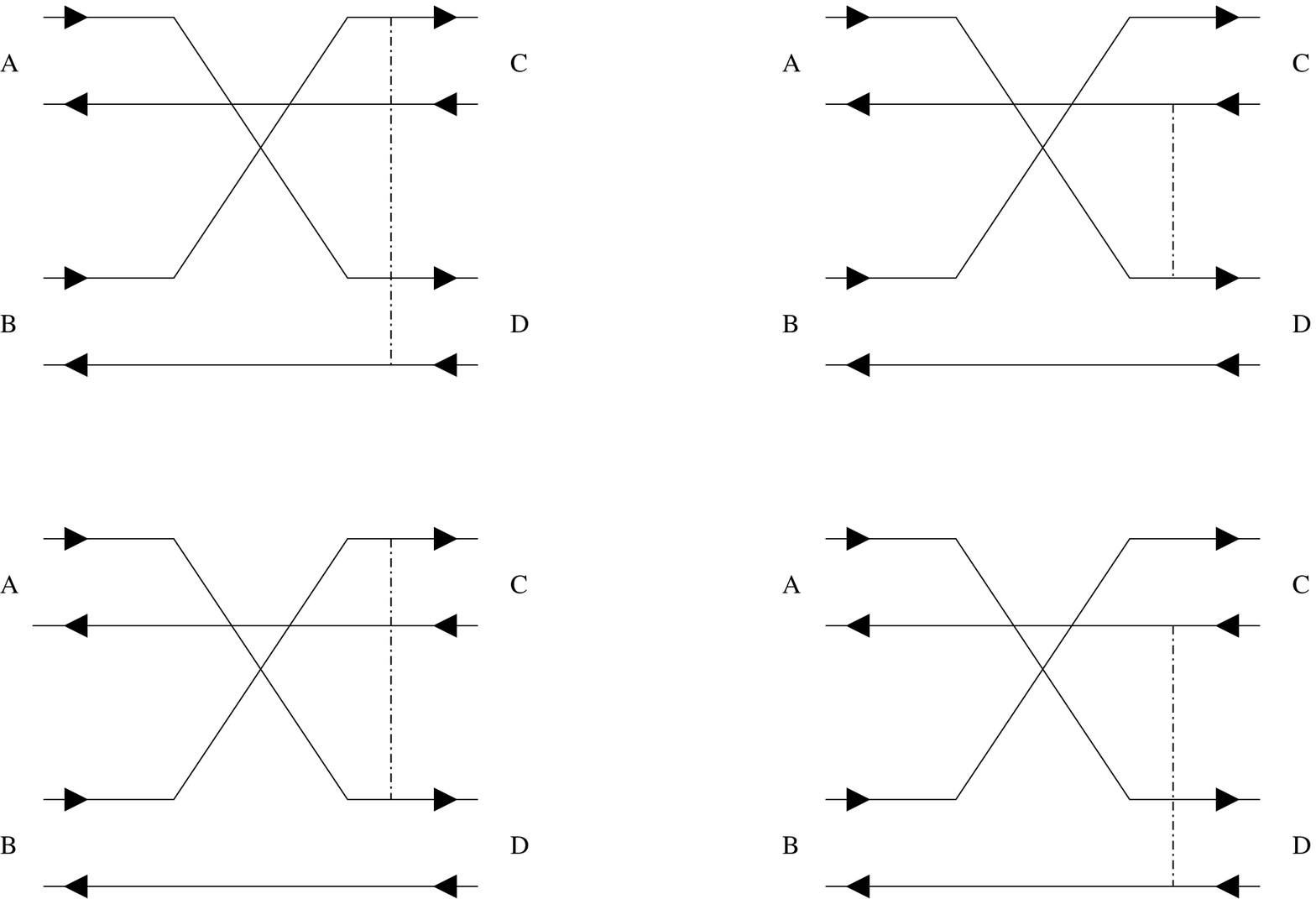}
\caption{Scattering in the post form. Solid lines with triangles right (left) 
represent quarks (antiquarks). Dot-dashed lines indicate interactions.}
\label{fig2}
\end{figure}

\newpage
\begin{figure}[htbp]
  \centering
    \includegraphics[scale=0.6]{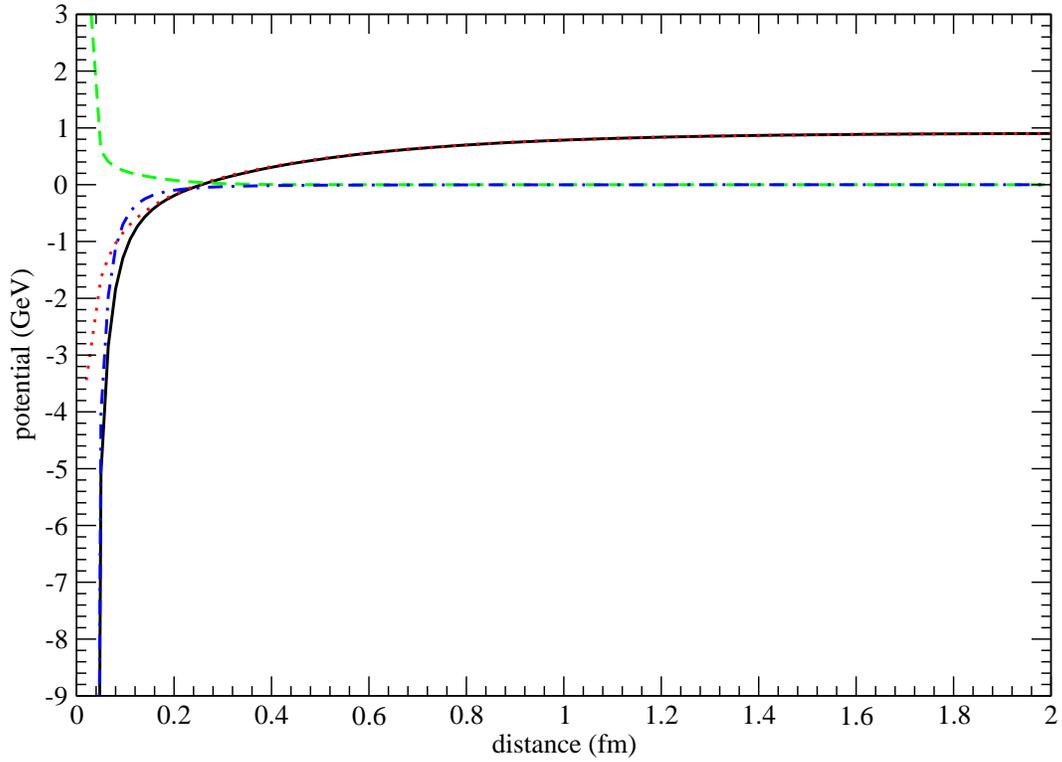}
\caption{Potentials as functions of the distance between the quark and the 
antiquark. 
The spin-independent potential, the spin-spin interaction, and the tensor
interaction are shown by the dotted, dashed, and 
dot-dashed curves, respectively. The solid curve indicates the sum of the
spin-independent potential, the spin-spin interaction, and the tensor
interaction.}
\label{fig3}
\end{figure}

\newpage
\begin{figure}[htbp]
  \centering
    \includegraphics[scale=0.6]{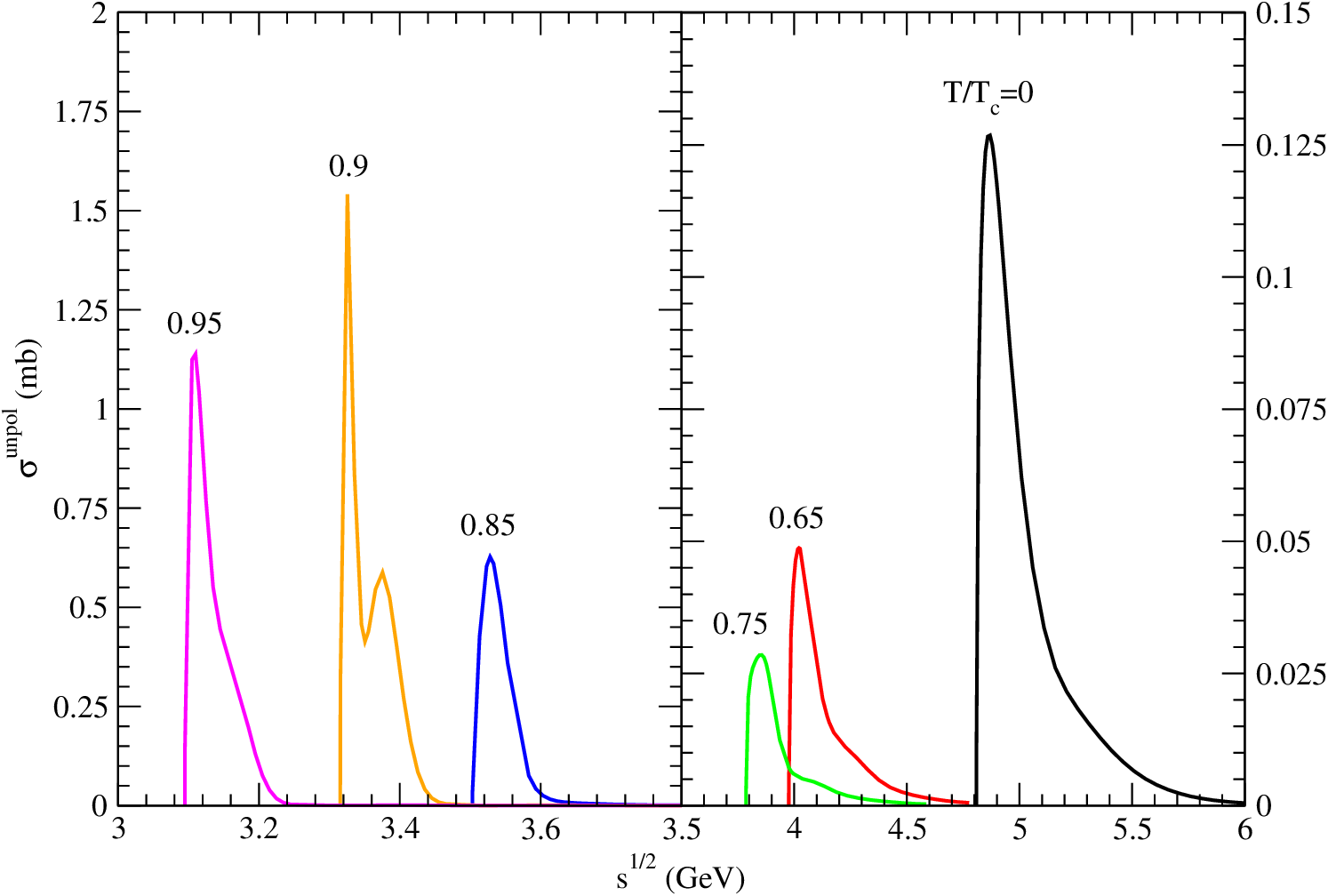}
\caption{Cross sections for $D\bar{D} \rightarrow \rho \psi(4040)$
at various temperatures.}
\label{fig4}
\end{figure}

\newpage
\begin{figure}[htbp]
  \centering
    \includegraphics[scale=0.6]{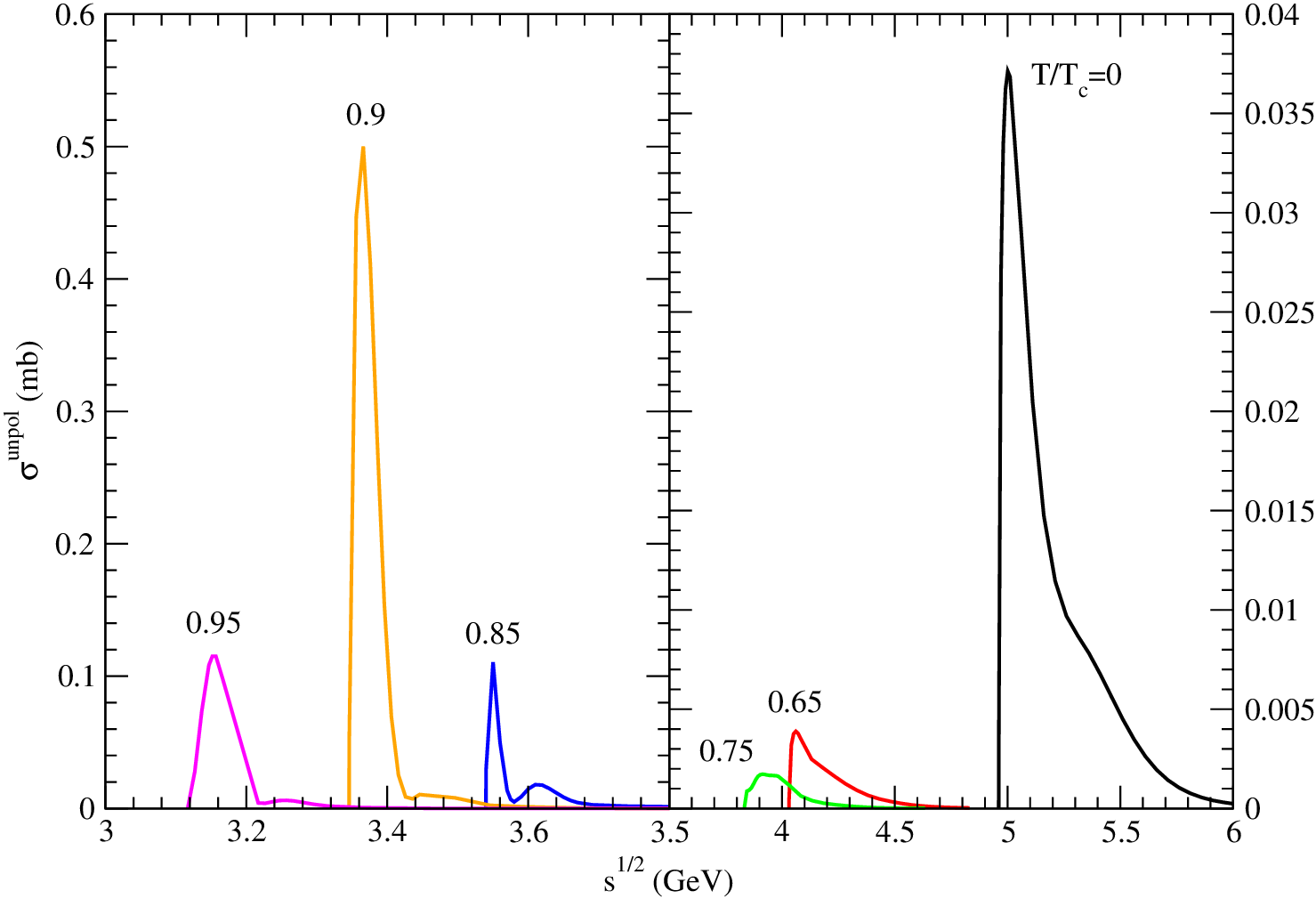}
\caption{Cross sections for $D\bar{D} \rightarrow \rho \psi(4160)$
at various temperatures.}
\label{fig5}
\end{figure}

\newpage
\begin{figure}[htbp]
  \centering
    \includegraphics[scale=0.6]{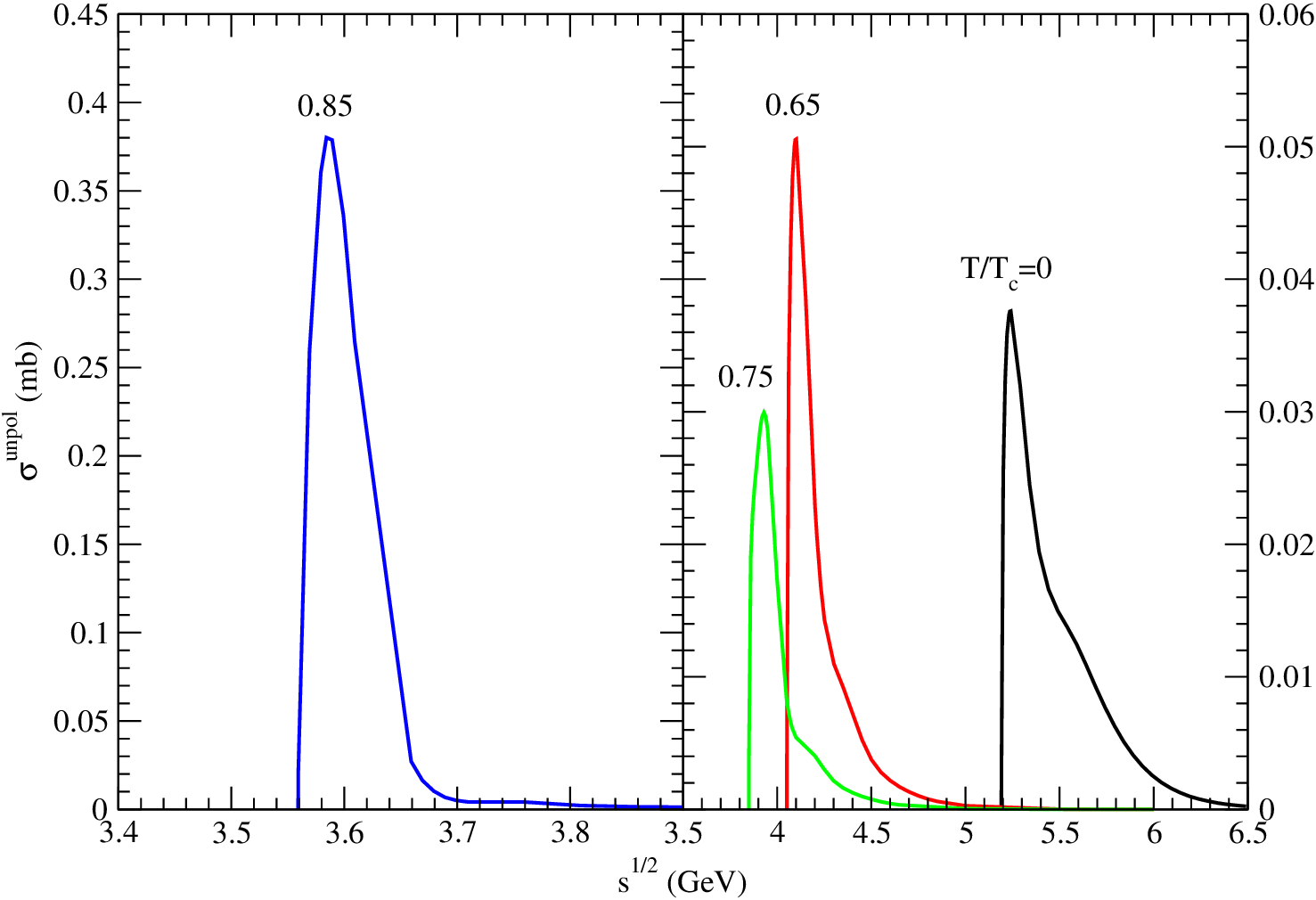}
\caption{Cross sections for $D\bar{D} \rightarrow \rho \psi(4415)$
at various temperatures.}
\label{fig6}
\end{figure}

\newpage
\begin{figure}[htbp]
  \centering
    \includegraphics[scale=0.6]{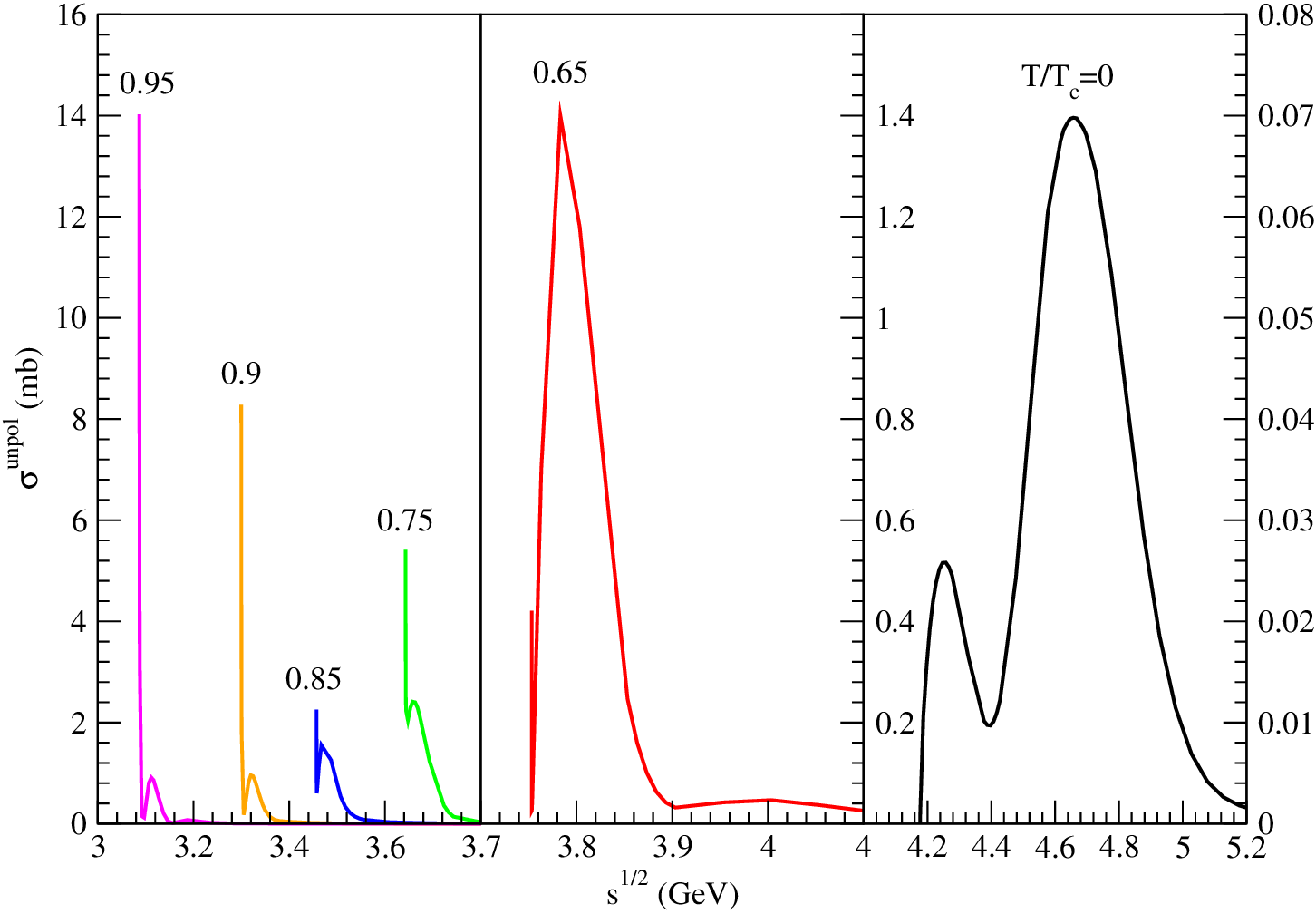}
\caption{Cross sections for $D\bar{D}^{\ast} \rightarrow \pi \psi(4040)$ 
at various temperatures.}
\label{fig7}
\end{figure}

\newpage
\begin{figure}[htbp]
  \centering
    \includegraphics[scale=0.6]{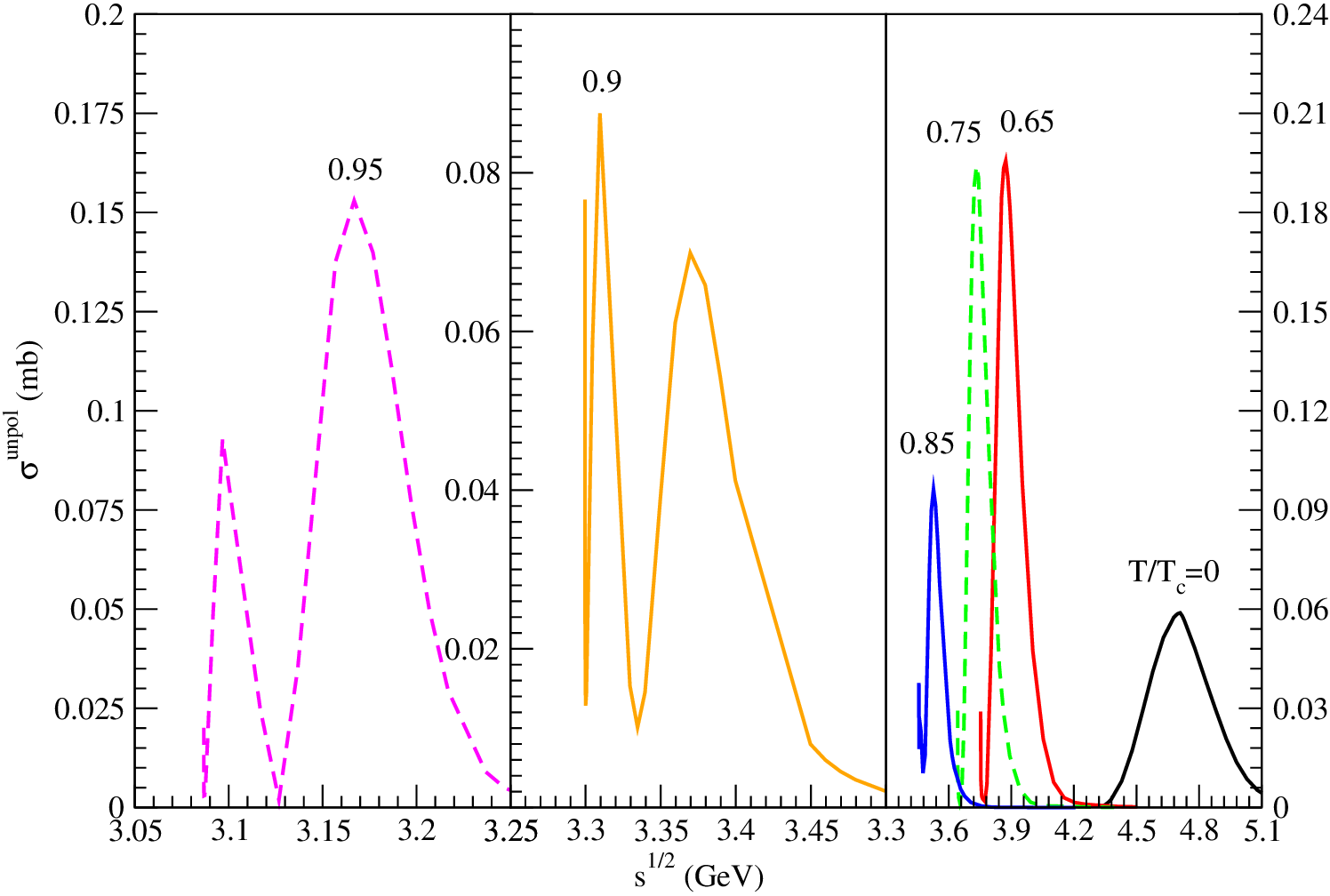}
\caption{Cross sections for $D\bar{D}^{\ast} \rightarrow \pi \psi(4160)$ 
at various temperatures.}
\label{fig8}
\end{figure}

\newpage
\begin{figure}[htbp]
  \centering
    \includegraphics[scale=0.6]{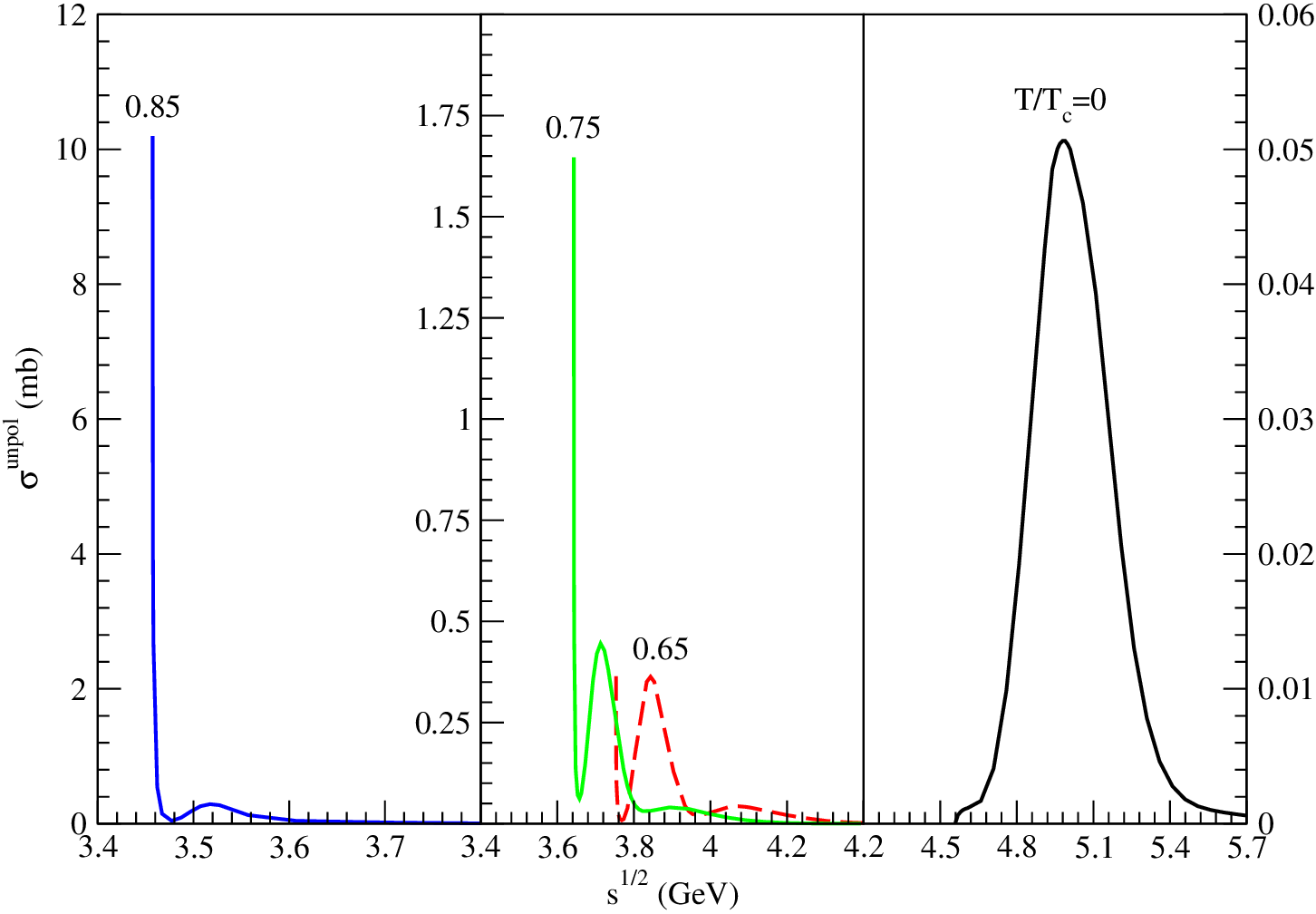}
\caption{Cross sections for $D\bar{D}^{\ast} \rightarrow \pi \psi(4415)$ 
at various temperatures.}
\label{fig9}
\end{figure}

\newpage
\begin{figure}[htbp]
  \centering
    \includegraphics[scale=0.6]{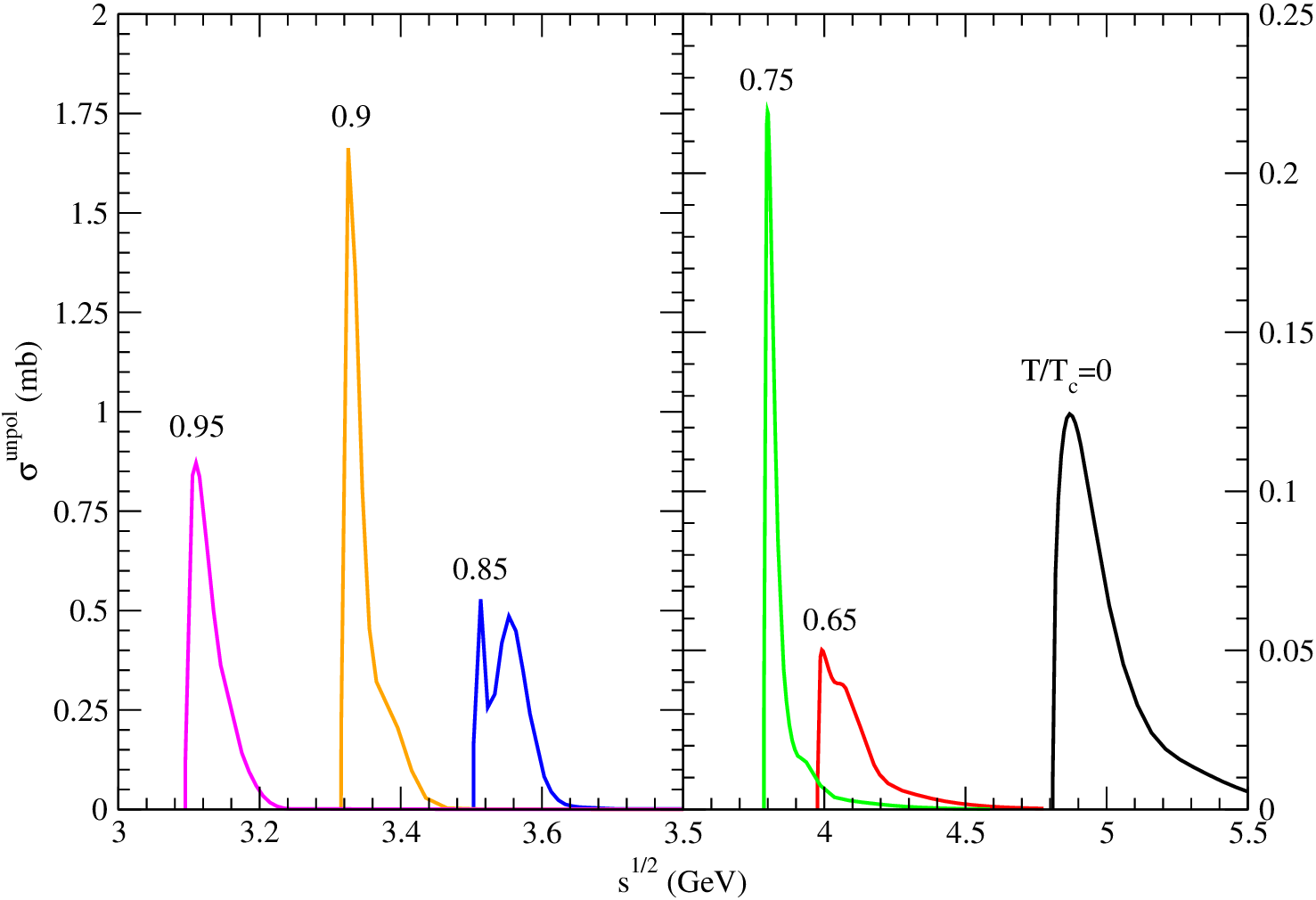}
\caption{Cross sections for $D\bar{D}^{\ast} \rightarrow \rho \psi(4040)$ 
at various temperatures.}
\label{fig10}
\end{figure}

\newpage
\begin{figure}[htbp]
  \centering
    \includegraphics[scale=0.6]{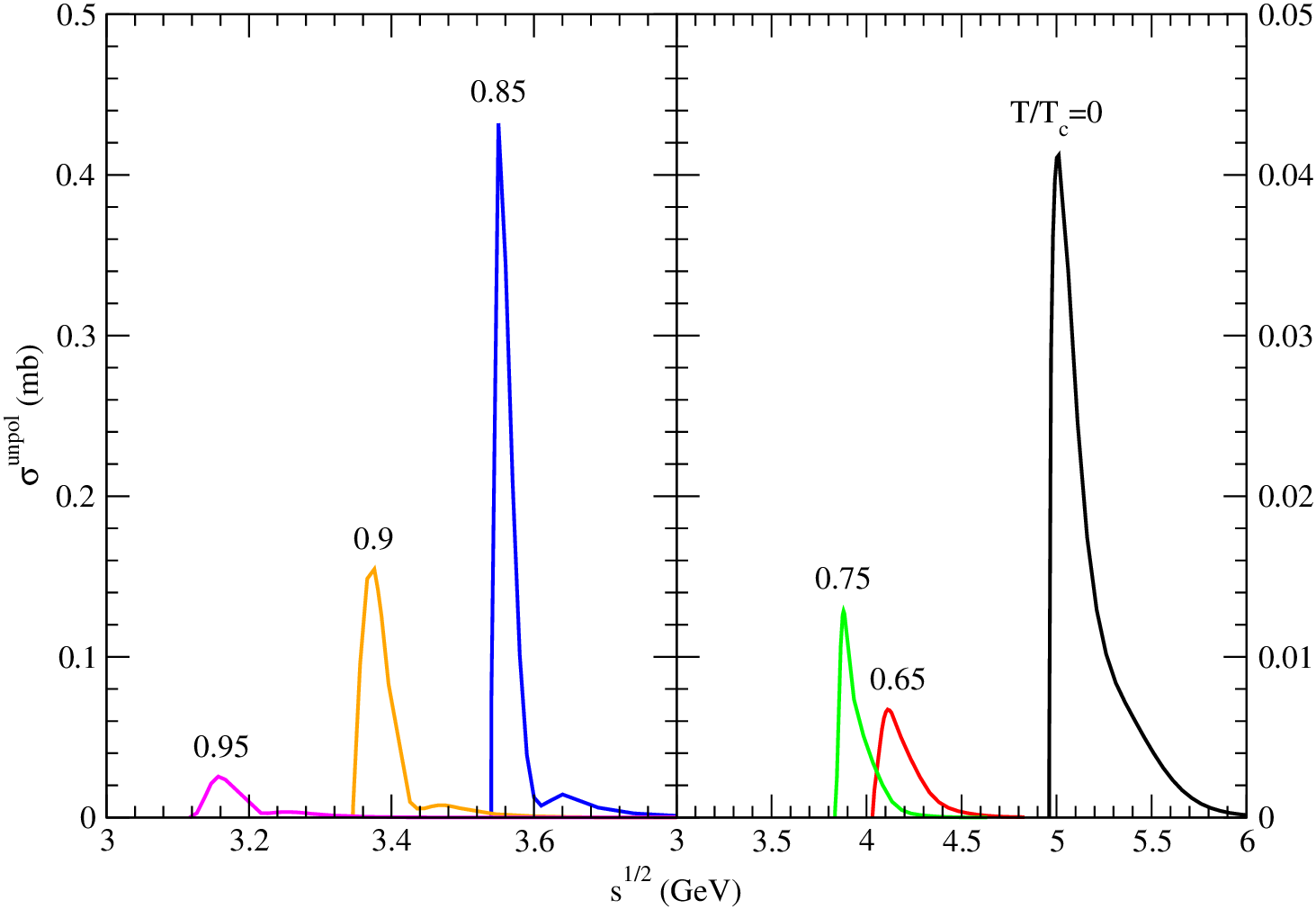}
\caption{Cross sections for $D\bar{D}^{\ast} \rightarrow \rho \psi(4160)$ 
at various temperatures.}
\label{fig11}
\end{figure}

\newpage
\begin{figure}[htbp]
  \centering
    \includegraphics[scale=0.6]{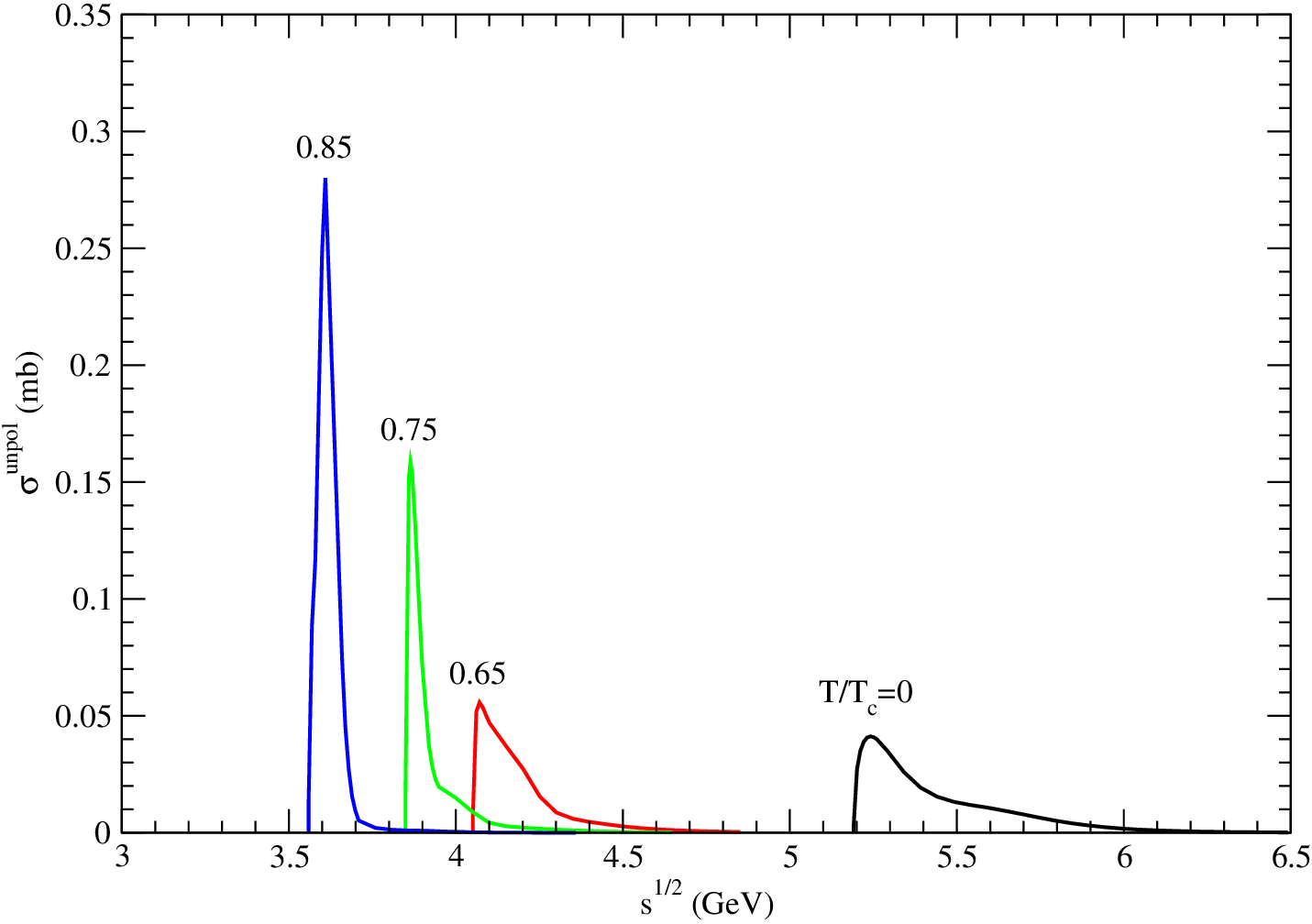}
\caption{Cross sections for $D\bar{D}^{\ast} \rightarrow \rho \psi(4415)$ 
at various temperatures.}
\label{fig12}
\end{figure}

\newpage
\begin{figure}[htbp]
  \centering
    \includegraphics[scale=0.6]{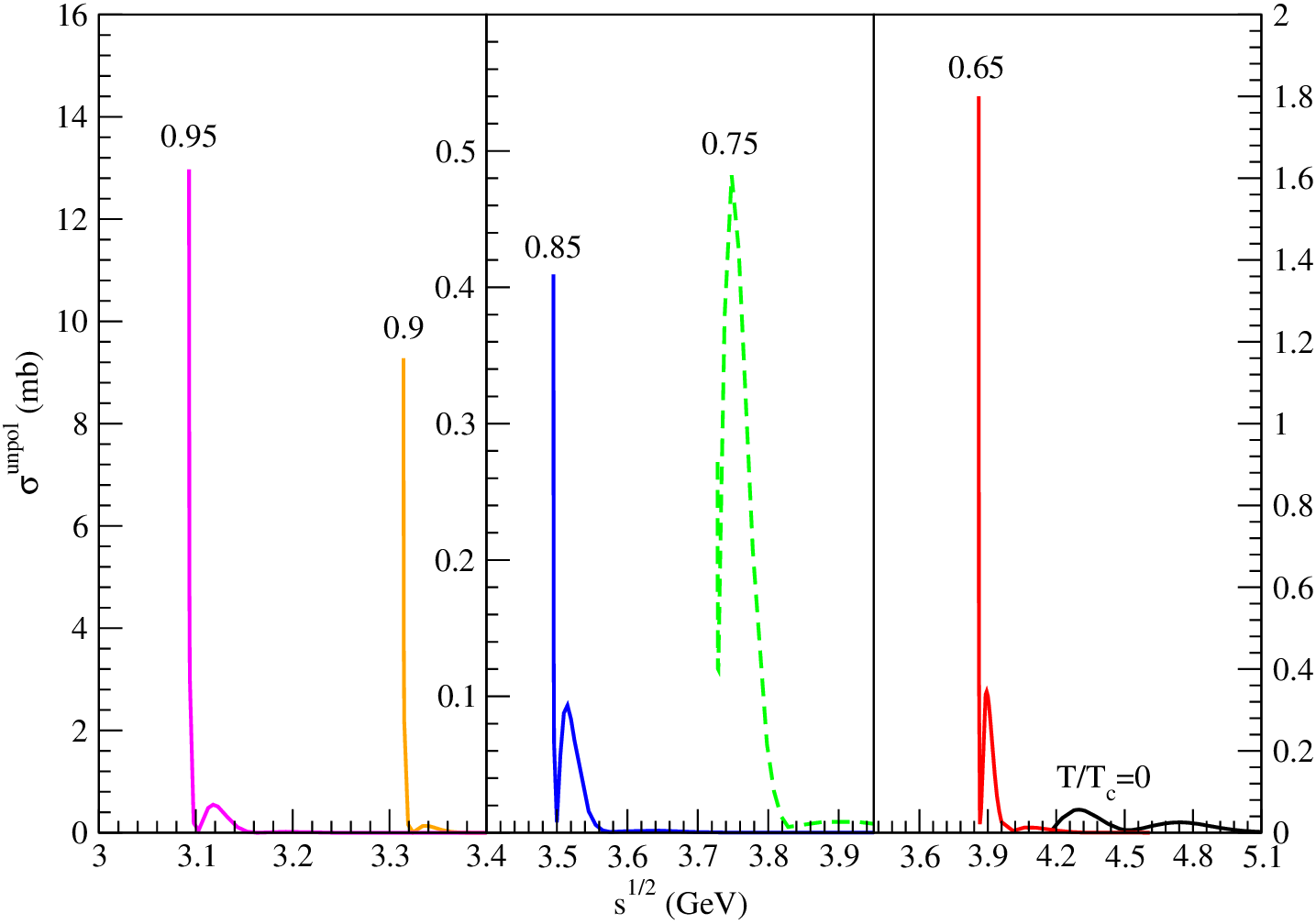}
\caption{Cross sections for $D^{\ast}\bar{D}^{\ast} \rightarrow \pi \psi(4040)$
at various temperatures.}
\label{fig13}
\end{figure}

\newpage
\begin{figure}[htbp]
  \centering
    \includegraphics[scale=0.6]{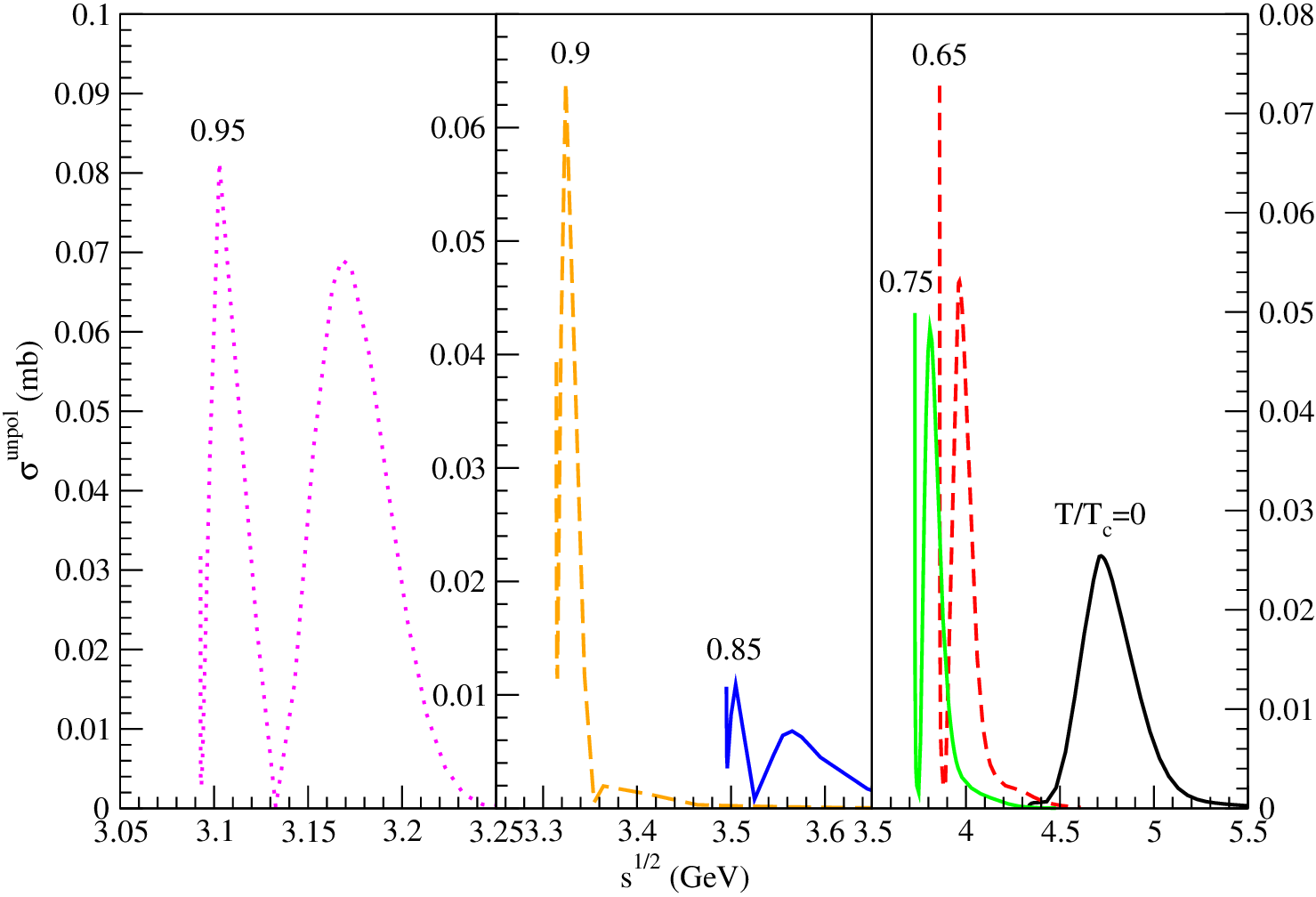}
\caption{Cross sections for $D^{\ast}\bar{D}^{\ast} \rightarrow \pi \psi(4160)$
at various temperatures.}
\label{fig14}
\end{figure}

\newpage
\begin{figure}[htbp]
  \centering
    \includegraphics[scale=0.6]{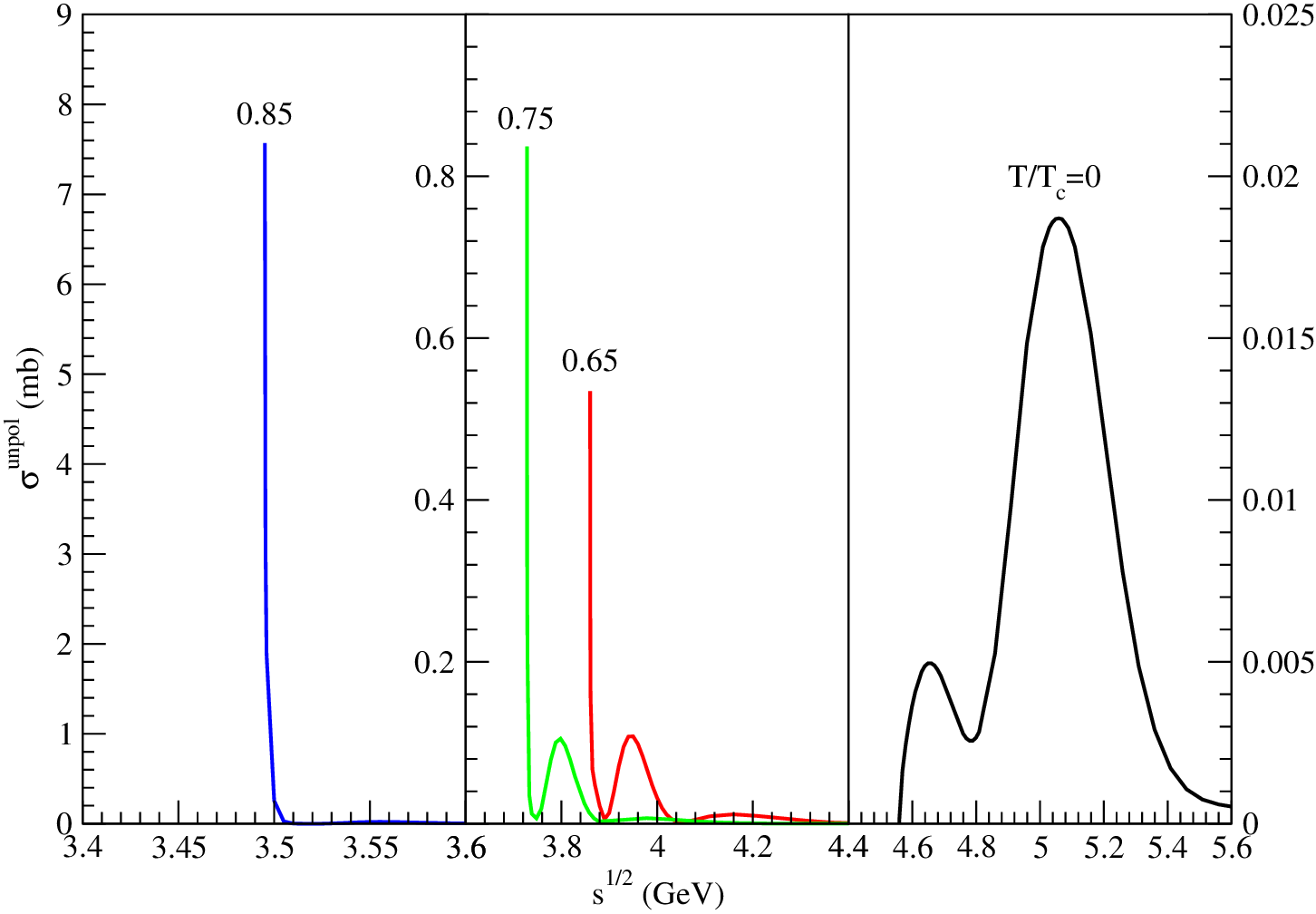}
\caption{Cross sections for $D^{\ast}\bar{D}^{\ast} \rightarrow \pi \psi(4415)$
at various temperatures.}
\label{fig15}
\end{figure}

\newpage
\begin{figure}[htbp]
  \centering
    \includegraphics[scale=0.6]{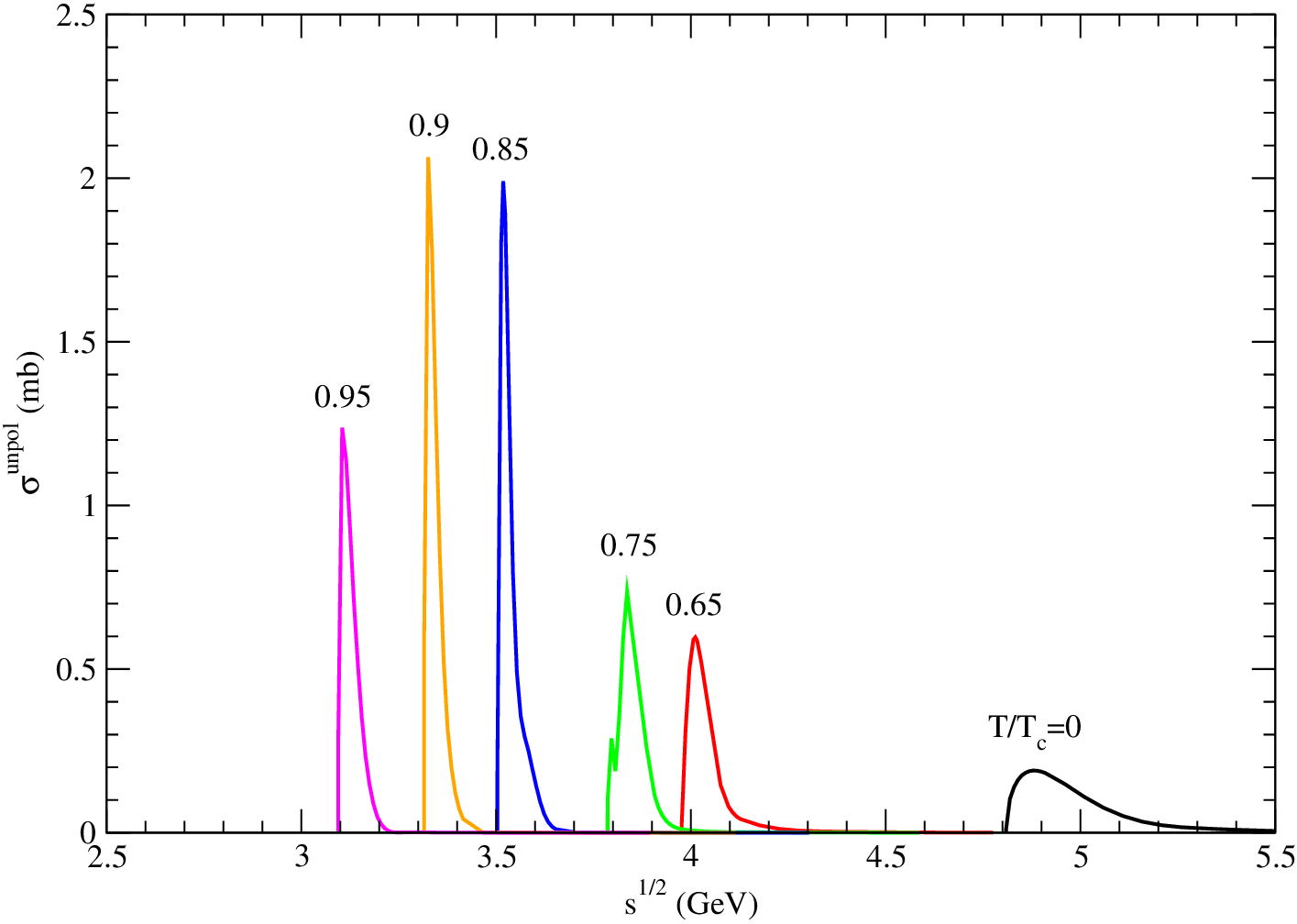}
\caption{Cross sections for $D^{\ast}\bar{D}^{\ast} \rightarrow \rho 
\psi(4040)$ at various temperatures.}
\label{fig16}
\end{figure}

\newpage
\begin{figure}[htbp]
  \centering
    \includegraphics[scale=0.6]{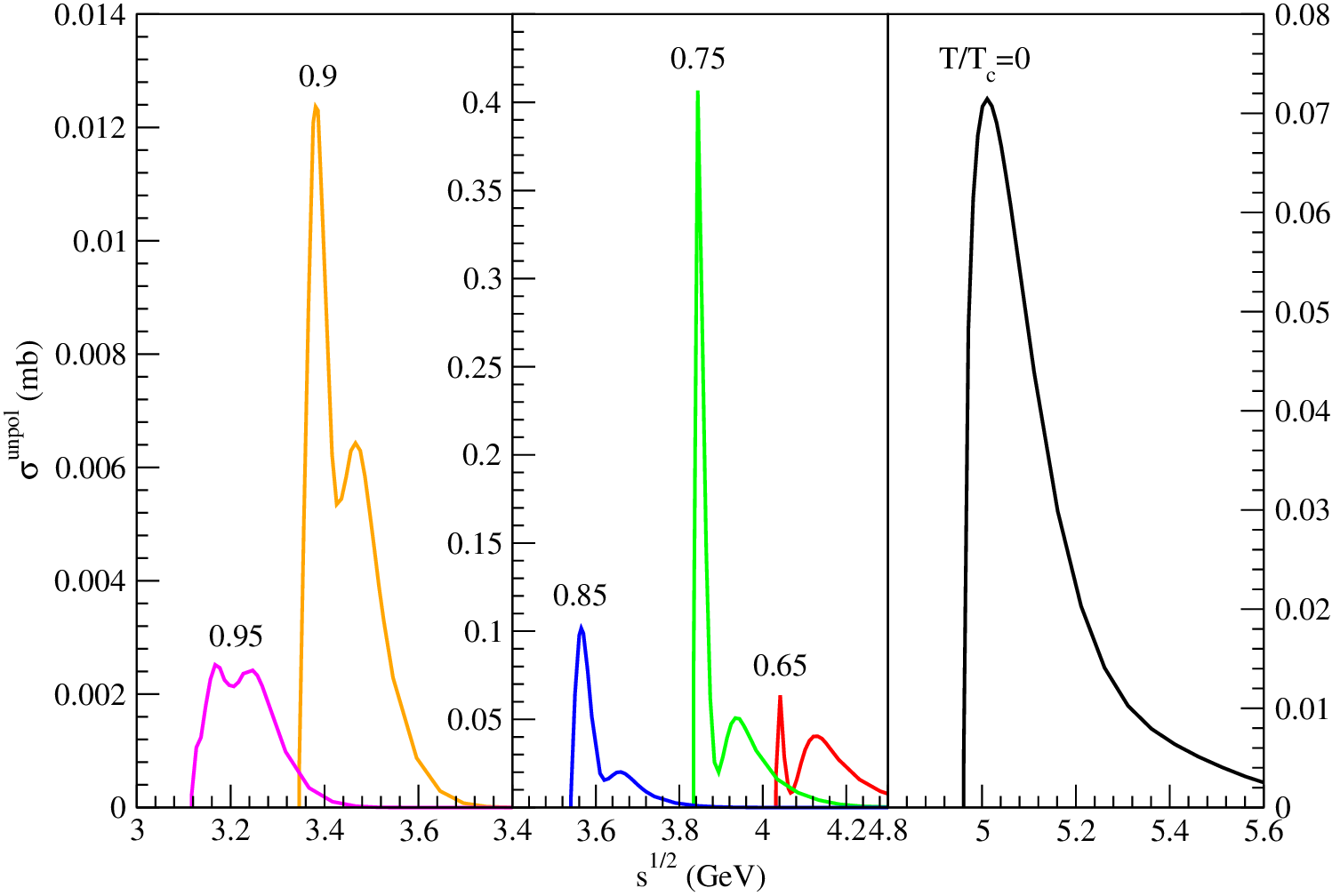}
\caption{Cross sections for $D^{\ast}\bar{D}^{\ast} \rightarrow \rho
\psi(4160)$ at various temperatures.}
\label{fig17}
\end{figure}

\newpage
\begin{figure}[htbp]
  \centering
    \includegraphics[scale=0.6]{dadarhopsi4415.eps}
\caption{Cross sections for $D^{\ast}\bar{D}^{\ast} \rightarrow \rho
\psi(4415)$ at various temperatures.}
\label{fig18}
\end{figure}

\newpage
\begin{figure}[htbp]
  \centering
    \includegraphics[scale=0.6]{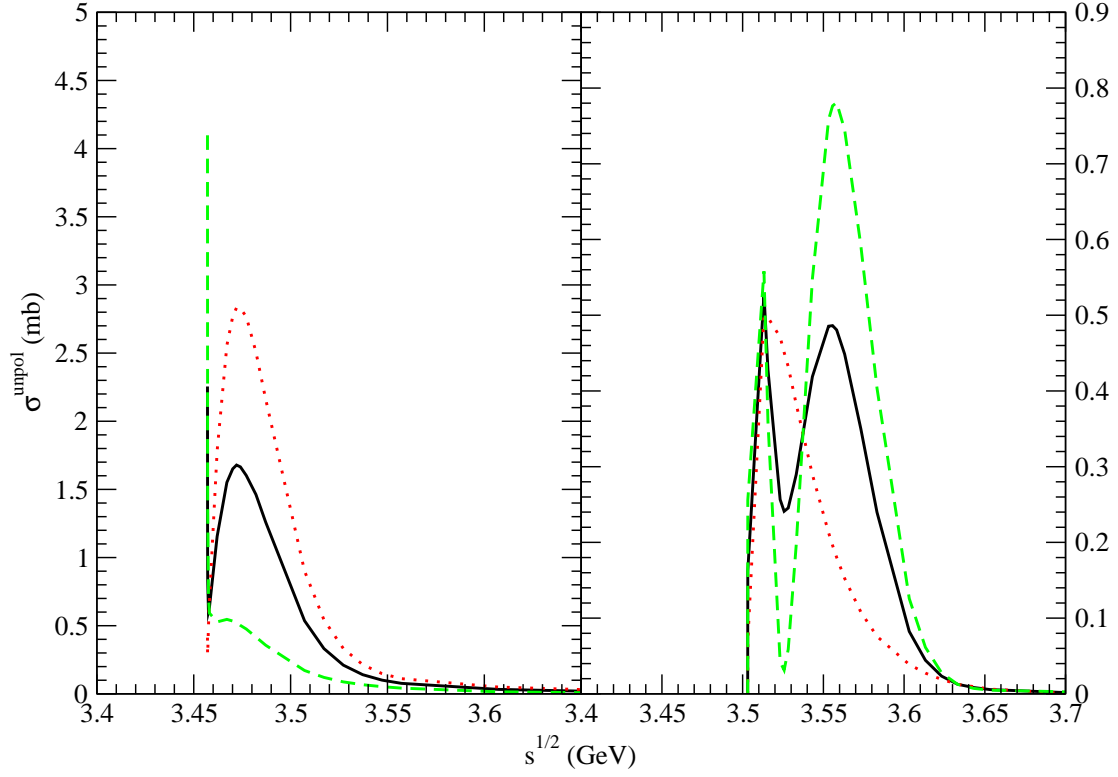}
\caption{Cross sections for $D\bar{D}^* \rightarrow \pi \psi(4040)$ in the 
left panel and for $D\bar{D}^* \rightarrow \rho \psi(4040)$ in the right panel
at $T/T_{\rm c}=0.85$. $\sigma^{\rm unpol}$, $\sigma_{\rm unpol}^{\rm prior}$,
and $\sigma_{\rm unpol}^{\rm post}$ are shown by the solid, dotted, and dashed
curves, respectively.}
\label{fig19}
\end{figure}

\newpage
\begin{figure}[htbp]
  \centering
    \includegraphics[scale=0.6]{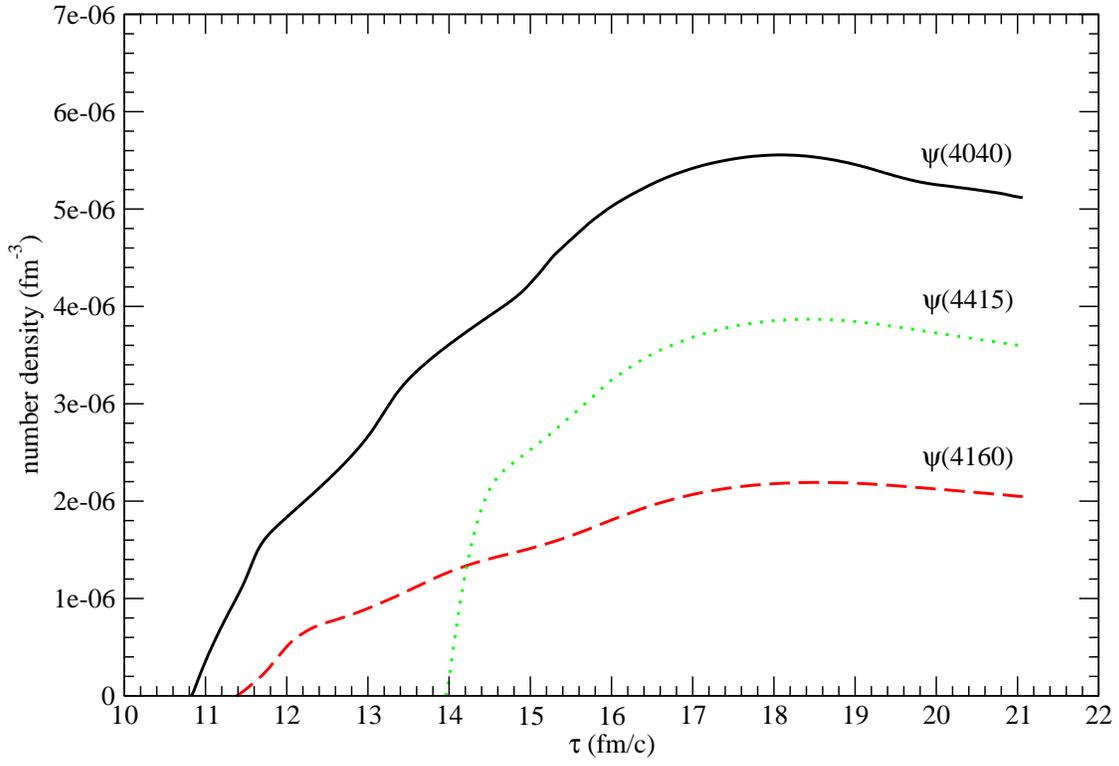}
\caption{Number densities as functions of $\tau$ at $r=0$ fm.}
\label{fig20}
\end{figure}

\newpage
\begin{figure}[htbp]
  \centering
    \includegraphics[scale=0.6]{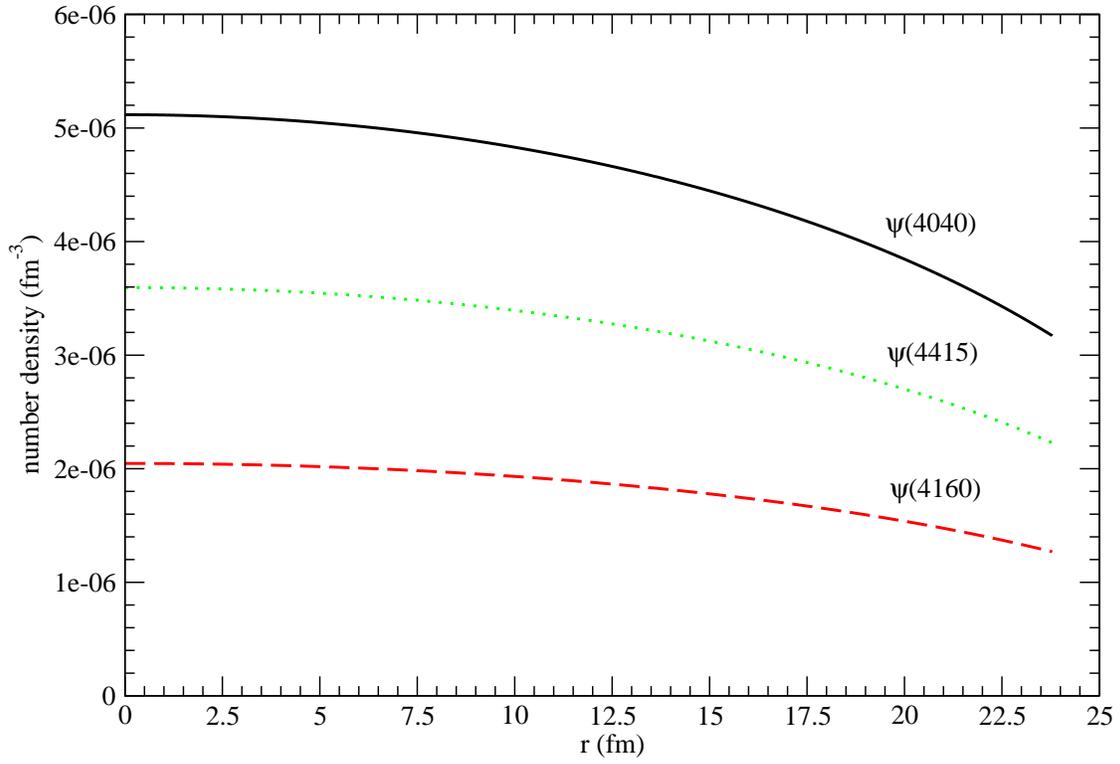}
\caption{Number densities as functions of $r$ at kinetic freeze-out.}
\label{fig21}
\end{figure}

\newpage
\begin{table*}[htbp]
\caption{\label{table1}Values of the parameters in Eq. (29) for $D\bar{D} \to 
\rho \psi(4040)$, $\rho \psi(4160)$, and $\rho \psi(4415)$. $a_1$ and $a_2$ are
in units of millibarns; $b_1$, $b_2$, $d_0$, and $\sqrt{s_{\rm z}}$ are
in units of GeV; $c_1$ and $c_2$ are dimensionless.}
\tabcolsep=5pt
\begin{tabular}{cccccccccc}
  \hline
  \hline
reaction & $T/T_{\rm c} $ & $a_1$ & $b_1$ & $c_1$ & $a_2$ & $b_2$ & $c_2$ &
$d_0$ & $\sqrt{s_{\rm z}} $\\
\hline
 $D\bar{D}\rightarrow\rho\psi(4040)$
   & 0     & 0.08  & 0.05  & 0.69  & 0.05 & 0.07  & 0.37  & 0.06  & 5.82\\
   & 0.65  & 0.04  & 0.04  & 0.88  & 0.01 & 0.09  & 0.18  & 0.045 & 4.79\\
   & 0.75  & 0.023 & 0.03  & 0.44  & 0.01 & 0.08  & 2.62  & 0.07  & 4.59\\
   & 0.85  & 0.45  & 0.03  & 2.59  & 0.28 & 0.01  & 0.62  & 0.025 & 3.65\\
   & 0.9   & 0.53  & 0.063 & 10.09 & 1.57 & 0.006 & 0.54  & 0.06  & 3.45\\
   & 0.95  & 0.76  & 0.01  & 0.85  & 0.42 & 0.02  & 0.32  & 0.015 & 3.23\\
  \hline
 $D\bar{D}\rightarrow\rho\psi(4160)$
   & 0     & 0.03   & 0.04  & 0.67 & 0.01   & 0.11  & 0.25 & 0.04   & 5.93 \\
   & 0.65  & 0.0028 & 0.05  & 0.36 & 0.0013 & 0.02  & 1.29 & 0.03   & 4.8 \\
   & 0.75  & 0.0018 & 0.1   & 1.69 & 0.0007 & 0.007 & 0.5  & 0.08   & 4.54 \\
   & 0.85  & 0.08   & 0.01  & 4.45 & 0.03   & 0.01  & 0.01 & 0.01   & 3.81 \\
   & 0.9   & 0.33   & 0.02  & 2.36 & 0.22   & 0.01  & 0.43 & 0.02   & 3.52 \\
   & 0.95  & 0.1    & 0.037 & 4    & 0.02   & 0.02  & 0.9  & 0.035  & 3.35 \\
  \hline
 $D\bar{D}\rightarrow\rho\psi(4415)$
   & 0     & 0.04  & 0.05 & 0.52 & 0.01  & 0.4  & 7.25 & 0.05   & 6.39 \\
   & 0.65  & 0.03  & 0.04 & 0.86 & 0.02  & 0.05 & 0.35 & 0.05   & 4.89 \\
   & 0.75  & 0.024 & 0.04 & 0.44 & 0.011 & 0.09 & 7.61 & 0.08   & 4.7 \\
   & 0.85  & 0.3   & 0.03 & 2    & 0.13  & 0.01 & 0.52 & 0.025  & 3.77 \\
  \hline
  \hline
\end{tabular}
\end{table*}
\begin{table*}[htbp]
\caption{\label{table2}Values of the parameters in Eqs. (29) and (30) for 
$D\bar{D}^* \to \pi \psi(4040)$, $\pi \psi(4160)$, and $\pi \psi(4415)$. 
$a_1$ and $a_2$ are
in units of millibarns; $b_1$, $b_2$, $d_0$, and $\sqrt{s_{\rm z}}$ are
in units of GeV; $c_1$ and $c_2$ are dimensionless.}
\tabcolsep=5pt
\begin{tabular}{cccccccccc}
  \hline
  \hline
reaction & $T/T_{\rm c} $ & $a_1$ & $b_1$ & $c_1$ & $a_2$ & $b_2$ & $c_2$ &
$d_0$ & $\sqrt{s_{\rm z}} $\\
\hline
 $D\bar{D}^{\ast}\rightarrow\pi\psi(4040)$
   & 0     & 0.08 & 0.49  & 16.02 & 0.022 & 0.061 & 0.61 & 0.48   & 5.37 \\
   & 0.65  & 1.15 & 0.04  & 2.81  & 0.14  & 0.08  & 1.31 & 0.05   & 4.22 \\
   & 0.75  & 1.27 & 0.032 & 2.32  & 0.3   & 0.026 & 0.64 & 0.03   & 4.03 \\
   & 0.85  & 0.96 & 0.026 & 1.97  & 0.07  & 0.054 & 0.59 & 0.03   & 3.75 \\
   & 0.9   & 1.09 & 0.028 & 4     & 0.11  & 0.001 & 0.6  & 0.025  & 3.54 \\
   & 0.95  & 3.42 & 0.029 & 6.73  & 0.81  & 0.001 & 0.62 & 0.03   & 3.28 \\
  \hline
  $D\bar{D}^{\ast}\rightarrow\pi\psi(4160)$
   & 0     & 0.048 & 0.35  & 3.65 & 0.003  & 0.5   & 0.62 & 0.38   & 5.39 \\
   & 0.65  & 0.52  & 0.123 & 6.38 & 0.0014 & 0.01  & 0.55 & 0.12   & 4.19 \\
   & 0.75  & 0.43  & 0.097 & 5.87 & 0.0008 & 0.006 & 0.53 & 0.1    & 3.99 \\
   & 0.85  & 0.2   & 0.086 & 8.26 & 0.012  & 0.029 & 0.93 & 0.08   & 3.74 \\
   & 0.9   & 0.146 & 0.08  & 4.16 & 0.065  & 0.015 & 2    & 0.07   & 3.54 \\
   & 0.95  & 0.83  & 0.077 & 7.54 & 0.45   & 0.016 & 2.44 & 0.08   & 3.27 \\
  \hline
  $D\bar{D}^{\ast}\rightarrow\pi\psi(4415)$
   & 0     & 0.045 & 0.43  & 7.82 & 0.002 & 0.29   & 0.51 & 0.43   & 5.74 \\
   & 0.65  & 0.91  & 0.091 & 5.96 & 0.014 & 0.004  & 0.45 & 0.09   & 4.34 \\
   & 0.75  & 0.9   & 0.074 & 5.1  & 0.046 & 0.006  & 0.42 & 0.08   & 4.15 \\
   & 0.85  & 0.61  & 0.065 & 6.34 & 0.232 & 0.0024 & 0.49 & 0.07   & 3.85 \\
  \hline
  \hline
\end{tabular}
\end{table*}
\begin{table*}[htbp]
\caption{\label{table3} The same as Table 1 except for $D\bar{D}^*$
reactions.}
\tabcolsep=5pt
\begin{tabular}{cccccccccc}
  \hline
  \hline
reaction & $T/T_{\rm c} $ & $a_1$ & $b_1$ & $c_1$ & $a_2$ & $b_2$ & $c_2$ &
$d_0$ & $\sqrt{s_{\rm z}} $\\
\hline
  $D\bar{D}^{\ast}\rightarrow\rho\psi(4040)$
   & 0     & 0.08 & 0.09 & 0.9  & 0.06 & 0.03  & 0.44 & 0.06   & 5.69 \\
   & 0.65  & 0.05 & 0.02 & 0.42 & 0.02 & 0.13  & 4.62 & 0.015  & 4.69 \\
   & 0.75  & 0.16 & 0.01 & 0.47 & 0.06 & 0.02  & 0.9  & 0.013  & 4.14 \\
   & 0.85  & 0.48 & 0.05 & 5.76 & 0.8  & 0.004 & 0.61 & 0.05   & 3.66 \\
   & 0.9   & 1.5  & 0.01 & 1.04 & 0.22 & 0.01  & 0.08 & 0.01   & 3.45 \\
   & 0.95  & 0.58 & 0.01 & 0.44 & 0.32 & 0.02  & 1.16 & 0.015  & 3.22 \\
  \hline
  $D\bar{D}^{\ast}\rightarrow\rho\psi(4160)$
   & 0     & 0.03  & 0.04  & 0.5  & 0.01  & 0.1  & 0.39 & 0.05   & 5.86 \\
   & 0.65  & 0.004 & 0.09  & 2.54 & 0.003 & 0.08 & 0.52 & 0.08   & 4.69 \\
   & 0.75  & 0.01  & 0.05  & 3.16 & 0.005 & 0.39 & 0.36 & 0.045  & 4.3 \\
   & 0.85  & 0.39  & 0.006 & 0.55 & 0.14  & 0.02 & 3.13 & 0.01   & 3.72 \\
   & 0.9   & 0.12  & 0.03  & 3.58 & 0.07  & 0.01 & 0.7  & 0.03   & 3.57 \\
   & 0.95  & 0.024 & 0.041 & 4.73 & 0.002 & 0.02 & 0.7  & 0.04   & 3.4 \\
  \hline
  $D\bar{D}^{\ast}\rightarrow\rho\psi(4415)$
   & 0     & 0.027 & 0.04  & 0.78 & 0.017 & 0.1   & 0.32 & 0.05   & 6.26 \\
   & 0.65  & 0.048 & 0.05  & 0.66 & 0.022 & 0.006 & 0.45 & 0.02   & 4.76 \\
   & 0.75  & 0.13  & 0.01  & 0.52 & 0.05  & 0.03  & 1.82 & 0.015  & 4.27 \\
   & 0.85  & 0.26  & 0.047 & 3.16 & 0.19  & 0.003 & 1.04 & 0.05   & 3.75 \\
  \hline
  \hline
\end{tabular}
\end{table*}
\begin{table*}[htbp]
\caption{\label{table4} The same as Table 2 except for $D^*\bar{D}^*$
reactions.}
\tabcolsep=5pt
\begin{tabular}{cccccccccc}
  \hline
  \hline
reaction & $T/T_{\rm c} $ & $a_1$ & $b_1$ & $c_1$ & $a_2$ & $b_2$ & $c_2$ &
$d_0$ & $\sqrt{s_{\rm z}} $\\
\hline
  $D^{\ast}\bar{D}^{\ast}\to\pi\psi(4040)$
   & 0     & 0.039  & 0.12   & 2.99  & 0.02   & 0.07   & 0.6  & 0.12  & 5.31\\
   & 0.65  & 0.18   & 0.044  & 4.34  & 0.006  & 0.0009 & 0.63 & 0.045 & 4.29\\
   & 0.75  & 0.17   & 0.03   & 2.44  & 0.007  & 0.03   & 0.63 & 0.03  & 4.07\\
   & 0.85  & 0.0515 & 0.0251 & 4.71  & 0.0021 & 0.0005 & 1.03 & 0.025 & 3.73\\
   & 0.9   & 0.154  & 0.0275 & 12.24 & 0.13   & 0.0011 & 0.69 & 0.025 & 3.38\\
   & 0.95  & 1.89   & 0.0287 & 8.12  & 0.68   & 0.001  & 0.6  & 0.03  & 3.16\\
  \hline
  $D^{\ast}\bar{D}^{\ast}\to\pi\psi(4160)$
   & 0     & 0.024  & 0.4   & 7.65  & 0.0006 & 0.05  & 0.52 & 0.39   & 5.51 \\
   & 0.65  & 0.075  & 0.116 & 7.05  & 0.0009 & 0.007 & 0.53 & 0.11   & 4.48 \\
   & 0.75  & 0.063  & 0.097 & 6.06  & 0.0005 & 0.005 & 0.54 & 0.09   & 4.26 \\
   & 0.85  & 0.0091 & 0.082 & 4.81  & 0.0035 & 0.014 & 2.01 & 0.07   & 3.93 \\
   & 0.9   & 0.048  & 0.014 & 3.09  & 0.004  & 0.12  & 0.48 & 0.01   & 3.55 \\
   & 0.95  & 0.37   & 0.076 & 15.79 & 0.32   & 0.017 & 2.42 & 0.075  & 3.24 \\
  \hline
  $D^{\ast}\bar{D}^{\ast}\to\pi\psi(4415)$
   & 0     & 0.019 & 0.51   & 16.47 & 0.0041 & 0.086  & 0.61  & 0.5   & 5.88\\
   & 0.65  & 0.16  & 0.09   & 12.74 & 0.0111 & 0.0076 & 0.72  & 0.09  & 4.4 \\
   & 0.75  & 0.108 & 0.074  & 6.96  & 0.0053 & 0.0023 & 0.36  & 0.07  & 4.19\\
   & 0.85  & 0.095 & 0.0015 & 0.57  & 0.03   & 0.064  & 12.41 & 0.001 & 3.51\\
  \hline
  \hline
\end{tabular}
\end{table*}
\begin{table*}[htbp]
\caption{\label{table5} The same as Table 1 except for $D^*\bar{D}^*$ 
reactions.}
\tabcolsep=5pt
\begin{tabular}{cccccccccc}
  \hline
  \hline
reaction & $T/T_{\rm c} $ & $a_1$ & $b_1$ & $c_1$ & $a_2$ & $b_2$ & $c_2$ &
$d_0$ & $\sqrt{s_{\rm z}} $\\
\hline
  $D^{\ast}\bar{D}^{\ast}\rightarrow\rho\psi(4040)$
   & 0     & 0.13 & 0.04  & 0.5  & 0.09 & 0.12  & 1.47 & 0.07   & 5.65 \\
   & 0.65  & 0.51 & 0.04  & 2.2  & 0.19 & 0.01  & 0.6  & 0.035  & 4.28 \\
   & 0.75  & 0.78 & 0.057 & 5.71 & 0.36 & 0.004 & 0.47 & 0.05   & 4 \\
   & 0.85  & 1.33 & 0.01  & 0.57 & 0.73 & 0.02  & 2.87 & 0.015  & 3.64 \\
   & 0.9   & 1.69 & 0.006 & 0.33 & 0.61 & 0.02  & 1.7  & 0.01   & 3.44 \\
   & 0.95  & 1.06 & 0.01  & 0.36 & 0.21 & 0.02  & 2.06 & 0.01   & 3.21 \\
  \hline
  $D^{\ast}\bar{D}^{\ast}\rightarrow\rho\psi(4160)$
   & 0     & 0.05   & 0.03   & 0.47 & 0.03   & 0.1   & 1.02  & 0.05  & 5.78 \\
   & 0.65  & 0.085  & 0.0042 & 0.58 & 0.04   & 0.108 & 4.45  & 0.01  & 4.5 \\
   & 0.75  & 0.32   & 0.01   & 1.76 & 0.09   & 0.01  & 0.04  & 0.01  & 4.14 \\
   & 0.85  & 0.07   & 0.03   & 3.36 & 0.05   & 0.01  & 0.67  & 0.025 & 3.84 \\
   & 0.9   & 0.008  & 0.03   & 6.22 & 0.0062 & 0.061 & 0.421 & 0.035 & 3.68 \\
   & 0.95  & 0.0017 & 0.104  & 2.34 & 0.0012 & 0.028 & 0.396 & 0.05  & 3.47 \\
  \hline
  $D^{\ast}\bar{D}^{\ast}\rightarrow\rho\psi(4415)$
   & 0     & 0.06 & 0.04  & 0.5  & 0.02 & 0.11  & 1.03 & 0.06   & 6.12 \\
   & 0.65  & 0.34 & 0.04  & 1.7  & 0.13 & 0.009 & 0.48 & 0.035  & 4.41 \\
   & 0.75  & 0.48 & 0.054 & 3.63 & 0.01 & 3.24  & 0.13 & 0.05   & 4.1 \\
   & 0.85  & 0.33 & 0.01  & 0.64 & 0.14 & 0.03  & 0.84 & 0.015  & 3.74 \\
  \hline
  \hline
\end{tabular}
\end{table*}

\end{document}